\documentclass[12pt]{article}
\pdfoutput=1
\usepackage{makeidx}
\makeindex
\usepackage[a4paper]{geometry}
\usepackage{jheppub,amsmath,amssymb,amsfonts,bm,amsxtra, mathrsfs,graphics,graphicx,amsthm,epsfig,ytableau,bm,longtable,float,color,tikz,mathtools,xfrac,footnote,rotating,lscape}
\usepackage{bookmark}

\usepackage{textgreek}
\usetikzlibrary{decorations.pathmorphing}
\usetikzlibrary{decorations.markings}
\usetikzlibrary{arrows, decorations.markings, calc, fadings, decorations.pathreplacing, patterns, decorations.pathmorphing, positioning}
\usepackage{tikz-cd}

\usetikzlibrary{positioning,shapes}
\usetikzlibrary{chains}
\usetikzlibrary{arrows,fit,decorations.pathreplacing}
\tikzstyle{every picture}+=[remember picture]
\tikzstyle{na} = [baseline=-.5ex]

\usepackage{empheq}
\usepackage{multirow}
\usepackage{booktabs}
\usepackage[american]{babel}

\usepackage[latin1]{inputenc}

\makeatletter
\newcommand{\vast}{\bBigg@{1}}
\newcommand{\Vast}{\bBigg@{5}}
\makeatother

\usepackage{hyperref}
\renewcommand{\arraystretch}{1.2}
\usepackage{setspace}

\numberwithin{equation}{section}

\newcommand{\cf}{\textit{cf.}}
\newcommand{\eg}{\textit{e.g.}}

\newcommand{\ie}{\textit{i.e.}}

\numberwithin{equation}{section}

\newcommand{\nn}{\nonumber}

\newcommand{\be}{\begin{equation}} \newcommand{\ee}{\end{equation}}
\newcommand{\bea}{\begin{equation} \begin{aligned}} \newcommand{\eea}{\end{aligned} \end{equation}}

\def\ISO{\mathrm{ISO}}
\def\SO{\mathrm{SO}}

\def\U{\mathrm{U}}
\def\SU{\mathrm{SU}}

\newcommand{\rd}{\mathrm{d}}
\newcommand{\e}{\mathrm{e}}

\newcommand{\Sg}{\Sigma_{\mathfrak{g}}}

\newcommand{\tsumI}{{\textstyle \sum_I}}

\DeclareMathOperator{\Tr}{Tr}

\newcommand{\cC}{\mathcal{C}}

\newcommand{\cF}{\mathcal{F}}

\newcommand{\cH}{\mathcal{H}}
\newcommand{\cI}{\mathcal{I}}

\newcommand{\cN}{\mathcal{N}}

\newcommand{\cS}{\mathcal{S}}

\newcommand{\cV}{\mathcal{V}}
\newcommand{\cW}{\mathcal{W}}

\newcommand{\bR}{\mathbb{R}}
\newcommand{\bZ}{\mathbb{Z}}
\newcommand{\fg}{\mathfrak{g}}
\newcommand{\fh}{\mathfrak{h}}
\newcommand{\fm}{\mathfrak{m}}

\newcommand{\fn}{\mathfrak{n}}

\newcommand{\fR}{\mathfrak{R}}

\usepackage[Symbol]{upgreek}
\usepackage{bm}

\usepackage{calligra}
\DeclareMathAlphabet{\mathcalligra}{T1}{calligra}{m}{n}

\title{Holographic microstate counting for \\ AdS$_4$ black holes in massive IIA supergravity}

\author[a,b]{Seyed Morteza Hosseini,}
\author[c]{Kiril Hristov}
\author[d]{and Achilleas Passias}

\affiliation[a]{Dipartimento di Fisica, Universit\`a di Milano-Bicocca, I-20126 Milano, Italy}
\affiliation[b]{INFN, sezione di Milano-Bicocca, I-20126 Milano, Italy}
\affiliation[c]{Institute for Nuclear Research and Nuclear Energy, Bulgarian Academy of Sciences, \\Tsarigradsko Chaussee 72, 1784 Sofia, Bulgaria}
\affiliation[d]{Department of Physics and Astronomy, Uppsala University, \\Box 516, SE-75120 Uppsala, Sweden}

\emailAdd{morteza.hosseini@mib.infn.it}
\emailAdd{khristov@inrne.bas.bg}
\emailAdd{achilleas.passias@physics.uu.se}

\preprint{UUITP-25/17}

\abstract{We derive the Bekenstein-Hawking entropy for a class of BPS black holes in the massive type IIA supergravity background AdS$_4 \times S^6$ from a microscopic counting of supersymmetric ground states in a holographically dual field theory. The counting is performed by evaluating the topologically twisted index of three-dimensional $\mathcal{N}=2$ Chern-Simons-matter gauge theories in the large $N$ limit. The $\mathcal{I}$-extremization principle is shown to match the attractor mechanism for the near-horizon geometries constructed in the four-dimensional dyonic $\mathcal{N}=2$ gauged supergravity, that arises as a consistent truncation of massive type IIA supergravity on $S^6$. In particular, our results prove that the imaginary part of the three-dimensional partition functions plays a crucial r\^ole in holography.} 

\begin{document}

\setcounter{tocdepth}{2}
\maketitle

%
%

\date{Dated: \today}




\section{Introduction}
\label{sec:introduction}

The topologically twisted index \cite{Benini:2015noa} is the partition function for three-dimensional
Chern-Simons-matter gauge theories preserving at least four real supercharges on $\Sigma_\fg \times S^1$,
with a partial topological A-twist on the genus $\fg$ Riemann surface $\Sigma_\fg$.\footnote{See \cite{Benini:2016hjo,Cabo-Bizet:2016ars,Closset:2016arn,Closset:2017zgf} for further developments.}
The index can be reduced to a matrix model exploiting the localization technique \cite{Benini:2015noa} and can be written as a contour integral,
\bea
Z (\fn, y) = \frac1{|\cW|} \; \sum_{\fm \,\in\, \Gamma_\fh} \; \oint_\cC Z_{\text{int}} (\fm, x;  \fn , y) \, ,
\eea
of a meromorphic differential form in variables $x = \e^{i u}$ ($u$ are the Coulomb branch parameters),
parameterizing the Cartan subgroup and subalgebra of the gauge group $G$
and summed over gauge magnetic fluxes $\fm$, living in the co-root lattice $\Gamma_\fh$ of $G$.
The index is a function of background magnetic fluxes $\fn_I$ and fugacities $y_I = \e^{i \Delta_I}$ for the global symmetries of the theory.

Large $N$ evaluation of the index for the three-dimensional ABJM theory \cite{Aharony:2008ug} was done
in \cite{Benini:2015eyy,Benini:2016rke} and successfully compared with the entropy of dyonic BPS black holes
in the M-theory background AdS$_4 \times S^7$.\footnote{See \cite{Cabo-Bizet:2017jsl} for a generalization of this setup to AdS$_4$ black holes with hyperbolic horizon. Recent attempts to compute the logarithmic corrections to the entropy of this class of black holes can be found in \cite{Liu:2017vll,Jeon:2017aif}.
It is of great importance to develop a new method to study the topologically twisted index beyond the leading large $N$ contribution.
It is naturally desirable to go beyond the numerical methods employed by \cite{Liu:2017vll}.}
Extending the results of \cite{Benini:2015eyy,Benini:2016rke}, the large $N$ limit of general three-dimensional
Chern-Simons-matter gauge theories with an M-theory or a massive type IIA dual was studied in \cite{Hosseini:2016tor,Hosseini:2016ume}.
For the special class of $\cN=2$ quiver gauge theories where the Chern-Simons levels
do not sum to zero the index has been shown to scale as $N^{5/3}$ in the large $N$ limit \cite{Hosseini:2016tor}, in agreement with a dual massive type IIA supergravity
construction \cite{Aharony:2010af,Petrini:2009ur,Lust:2009mb,Tomasiello:2010zz,Guarino:2015jca,Fluder:2015eoa,Pang:2015vna,Pang:2015rwd,Guarino:2016ynd,Guarino:2017eag,Guarino:2017pkw} (see also \cite{Araujo:2016jlx,Araujo:2017hvi}).

Motivated by the above results, we look at four-dimensional $\cN = 8$ supergravity with a dyonically gauged $\ISO(7) = \SO(7) \ltimes \bR^{7}$ gauge group
that arises as a consistent truncation of massive type IIA supergravity \cite{Romans:1985tz} on a six-sphere \cite{Guarino:2015vca,Cassani:2016ncu} and its further truncation to an $\cN = 2$ theory with an abelian gauge group $\mathbb{R} \times \U(1)^3$.
The electric and magnetic gauge couplings $(g,m)$ are identified with the $S^6$ inverse radius and the ten-dimensional Romans mass $\hat{F}_{(0)}$, respectively.
In particular, we analyze the supersymmetry conditions for black holes in AdS$_4 \times S^6$,
with deformed metrics on the $S^6$, in the presence of non-trivial scalar fields. We mainly focus on the near-horizon geometries which were also recently analyzed in \cite{Guarino:2017pkw}.
For our holographic purposes here we rederive these solutions in a different way and express the scalars and geometric data
in terms of the conserved electromagnetic charges. For the sake of clarity we focus primarily on the case of three magnetic
charges $\fn_j$ $(j=1,2,3)$ (with one constraint relating them) and equal electric charges $q_j=q\, ,\forall j=1,2,3$ with the possibility for different
horizon geometries of the form AdS$_2 \times \Sigma_\fg$.

The particular model we analyze corresponds to the $\cN=2$ truncation of the $\cN=8$ theory \cite{Guarino:2015vca}
coupled to three vector multiplets ($n_{\rm V}=3$) and the universal hypermultiplet ($n_{\rm H}=1$) \cite{Guarino:2017pkw}.
We will call this model the \emph{dyonic STU model}. The route that we take to constructing the near-horizon geometries
is based on a supersymmetry preserving version of the Higgs mechanism worked out in \cite{Hristov:2010eu} for the case of
$\cN=2$ gauged supergravity. This allows us to truncate away in a BPS preserving way a full massive vector multiplet
(made from the merging of the massless hypermultiplet and one of the three massless vector multiplets)
that forms after the spontaneous breaking of one of the gauge symmetries (corresponding to the $\mathbb{R}$ in $\mathbb{R} \times \U(1)^3$).
The remaining massless $\cN=2$ gauged supergravity contains only two vector multiplets and is described by the prepotential
 \bea
  \label{intro:prepotantial}
  \cF \left( X^I \right) = - i \frac{3^{3/2}}{4} \left(1-\frac{i}{\sqrt{3}}\right)
  c^{1/3} \left( X^1 X^2 X^3 \right)^{2/3} \ ,
 \eea
where the dyonic gauge parameter is the ratio $c \equiv m/g$.

The leading contribution to the black hole entropy is proportional to the area of its event horizon and it reads
\bea
 \label{intro:BH:entropy}
 S_{\rm BH} = \frac{{\rm Area}}{4 G_{\rm N}} \, ,
\eea
where $G_{\rm N}$ is the Newton's gravitational constant.
The number of black hole microstates $d_{\rm micro}$ should then be given by
\bea
 \label{d:micro}
 d_{\rm micro} = \e^{S_{\rm BH}} \, .
\eea
The goal of the current work is to verify this formula by a direct counting in the dual boundary description
in terms of a topologically twisted Chern-Simons-matter gauge theory with level $k$
given by the quantized Romans mass, $m=\hat F_{(0)}=k / (2 \pi \ell_s)$.\footnote{$\ell_s$ is the string length.}

The superconformal field theory (SCFT) dual to the background AdS$_4 \times S^6$ arises as an $\cN = 2$ Chern-Simons
deformation (at level $k$) of the maximal $\cN=8$ super Yang-Mills (SYM) theory on the worldvolume of $N$ D$2$-branes \cite{Schwarz:2004yj,Guarino:2015jca}.
We will call this model the \emph{D2$_k$ theory}.
It has an adjoint vector multiplet (containing a real scalar and a complex fermion) with gauge group $\U(N)$ or $\SU(N)$
and three chiral multiplets $\phi_{j}$ $(j = 1, 2, 3)$ (containing a complex scalar and fermion).

To verify \eqref{d:micro} we evaluate the topologically twisted index for ${\rm D2}_k$.
The final result is a function of magnetic charges $\fn_j$ and chemical potentials $\Delta_j$ for the global symmetries of the theory.
Reducing down to $S^1$, the theory gives rise to supersymmetric quantum mechanics and
the partition function on $\Sigma_\fg \times S^1$ computes the Witten index of the $\cN=2$ quantum mechanical sigma model \cite{Hori:2014tda,Hwang:2014uwa,Cordova:2014oxa}.
This naturally leads to a renormalization group (RG) flow across dimensions, connecting the CFT$_3$ dual to asymptotic AdS$_4$
vacuum in the ultraviolet (UV) and the CFT$_1$ dual to the near-horizon AdS$_2 \times \Sigma_\fg$ geometry in the infrared (IR).
Along the RG flow, the UV superconformal R-symmetry of the three-dimensional theory generically mixes with the flavor symmetries
and, at the one-dimensional fixed point, it becomes a linear combination of the reference R-symmetry and a subgroup of the flavor symmetries.
The R-symmetry that sits in the $\mathfrak{su}(1,1|1)$ superconformal algebra in the IR is determined by extremizing the topologically twisted index,
whose value at the extremum is the regularized number of ground states.
This is the so-called $\cI$-\emph{extremization principle} proposed in \cite{Benini:2015eyy,Benini:2016rke}.

Let us state the main result of our paper. Upon extremizing $\cI (\Delta_j)$, at large $N$,
with respect to the chemical potentials $\Delta_j$ we show that its value at the extremum $\bar\Delta_j$ precisely reproduces the black hole entropy:
\bea
 \label{main_result}
 \cI ( \bar\Delta_j ) \equiv \log Z ( \bar\Delta_j ) - i \sum_{j=1}^{3} \bar\Delta_j q_j
 = S_{\rm BH} ( \fn_j , q_j ) \, .
\eea

Furthermore, during the course of our analysis we find a number of intriguing results, valid at large $N$.

First, we show that for a class of $\cN=2$ Chern-Simons-matter gauge theories where the Chern-Simons levels do not add up to zero,
the logarithm of the topologically twisted index can be written as\footnote{A similar relation between the three-sphere free energy $- \log Z_{S^3} \left( \Delta_I \right)$
and the anomaly coefficient $a \left( \Delta_I \right)$ was found in \cite{Fluder:2015eoa}.
}
\bea
 \label{intro:index:generic:c2d:a4d}
 \log Z \left( \fn_I, \Delta_I \right) =
 - \frac{ 3^{7/6} \pi}{5 \times 2^{10/3}}
 \left( 1 - \frac{i}{\sqrt{3}} \right) 
 \left( n N \right)^{1/3}
 \frac{c_r \left( \fn_I , \Delta_I \right)}{a \left( \Delta_I \right)^{1/3}} \, ,
\eea
where $n \equiv \sum_{a=1}^{|G|} k_a$.
Here $a \left( \Delta_I \right)$ is the trial conformal 't Hooft anomaly of a ``parent'' four-dimensional $\cN=1$ superconformal field theory on $S^2 \times T^2$
and $c_r \left( \fn_I , \Delta_I \right)$ is the trial right-moving central charge of the $\cN=(0,2)$ theory obtained by a twisted compactification on $S^2$ \cite{Benini:2012cz,Benini:2013cda,Hosseini:2016cyf}.
Here we use the chemical potentials $\Delta_I / \pi$ to parameterize a trial R-symmetry of the theory.
Details about this relation are given in the main text.

Given the interesting connection between the four-dimensional D3-brane theories and the three-dimensional D2$_k$ theories \eqref{intro:index:generic:c2d:a4d},
it would be intriguing to find the analogous relation on the supergravity side. In particular, one can expect a close connection
between the supergravity backgrounds discussed here and the black string solutions in five-dimensional
STU gauged supergravity found in \cite{Benini:2013cda}.

Secondly, we demonstrate another example of the conjecture originally posed in \cite{Hosseini:2016tor}:
\bea
 \label{intro:extr:attractor}
 - \log Z_{S^3} \left( \Delta_j \right) & \propto \cF \left( X^j \right) \, , \\
 \cI\text{-extremization} & = \text{attractor mechanism} \, ,
\eea
where $Z_{S^3} \left( \Delta_j \right)$ denotes the $S^3$ partition function for ${\rm D2}_k$,
depending on trial R-charges $\Delta_j$ \cite{Fluder:2015eoa}:
\bea
 \label{intro:S3 free energy}
 \log Z_{S^3} = - \frac{3^{13/6} \pi }{5 \times 2^{5/3}} \left( 1 - \frac{i}{\sqrt{3}} \right) k^{1/3} N^{5/3}
 \left( \Delta_1 \Delta_2 \Delta_3 \right)^{2/3} \, .
\eea
Quite remarkably, the above correspondence \eqref{intro:extr:attractor} holds true including the imaginary part of the $S^3$ partition function and the prepotential \eqref{intro:prepotantial}.

The remainder of this paper is arranged as follows.
In section \ref{sec:twisted:index} we shall give a short review of the topologically twisted index and its large $N$ limit.
We focus on a class of three-dimensional supersymmetric Chern-Simons-matter gauge theories arising from D$2$-branes
probing generic Calabi-Yau three-fold (CY$_3$) singularities in the presence of non-zero quantized Romans mass.
We also derive the formula \eqref{intro:index:generic:c2d:a4d}.
Then we move to evaluate the twisted index for the $\cN=2$ ${\rm D2}_k$ theory.
In section \ref{sec:dyonic sugra} we switch gears and discuss our supergravity
solutions dual to a topologically twisted deformation of the ${\rm D}2_k$ theory.
This section contains a brief overview of the four-dimensional $\cN=2$ dyonic STU model,
as constructed in \cite{Guarino:2017pkw}, and the supersymmetric conditions for the existence of black hole solutions. 
We then proceed to analyze in more detail the exact UV and IR limits of the general equations, recovering
the asymptotic AdS$_4$ and the near-horizon AdS$_2 \times \Sigma_\fg$ geometries.
We finish this section by commenting on the general existence of full BPS flows between the UV and IR solutions that we have. 
In section \ref{sec:index vs entropy} we compare the field theory and the supergravity results,
and we show that the $\cI$-extremization correctly reproduces the black hole entropy.
Finally we have appendix \ref{appendix} with all the fine details of the supergravity model
and solutions that are only sketched in section \ref{sec:dyonic sugra} for the sake of clarity. 

\paragraph*{Note added:} The counting of microstates for black holes with constant scalar fields
--- equal fluxes along the exact R-symmetry of three-dimensional SCFTs ---
and horizon topology AdS$_2 \times \Sigma_\fg$, $(\fg>1)$ has been recently considered in \cite{Azzurli:2017kxo}.
While we were completing this work, we became aware of \cite{Benini:2017oxt}
which we understand has overlap with the results presented here.

\section{The topologically twisted index}
\label{sec:twisted:index}

The topologically twisted index of a three-dimensional $\cN = 2$
supersymmetric gauge theory is defined as the partition function
of the theory placed on the background $\Sigma_\fg \times S^1$, with a partial
topological A-twist along the genus $\fg$ Riemann surface $\Sigma_\fg$,
in the presence of background magnetic fluxes $\fn_I$, parameterizing the twist,
and fugacities $y_I = \e^{i v_I}$ for the global symmetries \cite{Benini:2015noa}.
The index can be interpreted as a trace over a Hilbert space of states $\cH$ on $\Sigma_\fg$,
\begin{equation}
 Z(\fn_I, v_I) = \Tr\nolimits_{\cH} (-1)^{F} \e^{i \sum_I \Delta_I J_I} \e^{-\beta H} \, ,
\end{equation}
where $J_I$ are the generators of the flavor symmetries.
The Hamiltonian $H$ on $\Sigma_\fg$ explicitly depends on the flavor magnetic fluxes $\fn_I$ and the real masses $\sigma_I$.
Due to the supersymmetry algebra $Q^2 = H - \sigma_I J_I$ only states with $H=\sigma_I J_I$ contribute.
The index is a holomorphic function of $v_I$ with $v_I = \Delta_I + i \beta \sigma_I$.\footnote{$\beta$ is the radius of $S^1$.}
We also identify $\Delta_I$ with flavor flat connections.
Furthermore, we limit ourselves to the case of $\Sigma_\fg = S^2$ since the generalization to an arbitrary Riemann surface is straightforward, see section 6 of \cite{Benini:2016hjo}.

The topologically twisted index can be computed using supersymmetric localization and it is given by
a contour integral over the zero-mode gauge variables $x = \e^{i (A_t + i \beta \sigma)}$,
parameterizing the Cartan subgroup of the gauge group $G$, and it is summed over a lattice of
gauge magnetic charges $\fm$, living in the co-root lattice $\Gamma_\fh$ of $G$ (up to gauge transformations),
on $S^2$ \cite{Benini:2015noa}.
Here $A_t$ is a Wilson line on $S^1$ which runs over the maximal torus of $G$ and $\sigma$ is the real scalar
in the vector multiplet which runs over the corresponding Cartan subalgebra.
We focus on Chern-Simons quiver gauge theories with bi-fundamental and adjoint chiral multiplets
transforming in representations $\fR_I$ of $G$ and a number $|G|$ of $\U(N)$ gauge groups with equal Chern-Simons couplings $k_a = k$ $(a = 1, \ldots, |G|)$. 
Explicitly, the index can be written as\footnote{Supersymmetric localization chooses a particular contour of integration $\cC$
and the final result can be recast in terms of the Jeffrey-Kirwan residue.
We refer the reader to \cite{Benini:2015noa,Benini:2016hjo} for a thorough analysis of the contour of integration.}
\be
 \label{index:path integral}
 Z (\fn_I, y_I) = \frac1{|\cW|} \; \sum_{\fm \,\in\, \Gamma_\fh} \oint_\cC \
 \prod_{\mathrm{Cartan}} \left (\frac{\rd x}{2\pi i x}  x^{k \fm} \right ) \prod_{\alpha \in G} (1-x^\alpha) \ 
 \prod_I \prod_{\rho_I \in \fR_I} \bigg( \frac{x^{\rho_I/2} \, y_I^{1/2}}{1-x^{\rho_I} \, y_I} \bigg)^{\rho_I(\fm)- \fn_I  +1} \, ,
\ee
where the index $I$ runs over all matter fields in the theory. Given a weight $\rho_I$ of the representation $\fR_I$, we employed the notation $x^{\rho_I} = \e^{ i \rho_{I}(u) }$.
Here $\alpha$ denote the roots of the gauge group $G$ and $|\cW|$ is the order of the Weyl group of $G$.

In the next section we shall evaluate the index at large $N$ for real $v_I$, setting all real masses $\sigma_I$ to zero.
One can easily extend it to the complex plane employing holomorphy.

\subsection[The large \texorpdfstring{$N$}{N} limit of the index for a generic theory]{The large $\bm{N}$ limit of the index for a generic theory}
\label{ssec:largeN:limit:index}

We are interested in the large $N$ limit, $N \gg k_a$ with $\sum_{a=1}^{|G|}k_a \neq 0$, of the index for Chern-Simons-matter gauge theories with massive type IIA supergravity
duals $\mathrm{AdS}_4 \times \cS Y_5$ \cite{Aharony:2010af,Petrini:2009ur,Lust:2009mb,Tomasiello:2010zz,Guarino:2015jca,Fluder:2015eoa,Pang:2015vna,Pang:2015rwd,Guarino:2016ynd,Guarino:2017eag,Guarino:2017pkw}. Here $\cS Y_5$ denotes the suspension of a generic Sasaki-Einstein five-manifold $Y_5$.
These theories describe the dynamics of $N$ D$2$-branes probing a generic Calabi-Yau three-fold (CY$_3$) singularity in the presence of a non-vanishing quantized Romans mass $m$ \cite{Gaiotto:2009mv}.
At large $N$, it is natural to describe the eigenvalues $u^{(a)}$ in terms of their density $\rho(u)$ with the constraint $\int {\rm d}t \, \rho(t) = 1$ --- equivalently, we shall introduce a Lagrange multiplier $\mu$.
Taking the ansatz $u^{(a)} (t) = N^{1/3} \left( i t + v(t) \right)$ for the eigenvalue distribution \cite{Jafferis:2011zi,Fluder:2015eoa}, one can show that
the so-called \emph{long-range forces} cancel automatically if the conditions
\bea
 \label{BethePot:constraint:generic}
 \Tr R = |G| + \sum_I \left( \fn_I - 1 \right) = 0 \, , \qquad \qquad
 \Tr J = \pi |G| + \sum_I \left( \Delta_I - \pi \right) = 0 \, ,
\eea
are met by the Chern-Simons-matter gauge theories \cite{Hosseini:2016tor}. Here the trace is taken over bi-fundamental fermions and gauginos in the quiver.
For quivers with a four-dimensional ``parent'',\footnote{They describe D$3$-branes probing the same CY$_3$ singularity.}
these conditions are equivalent to the absence of anomalies for the R- and the global symmetries of the theory in four dimensions.
The above constraints \eqref{BethePot:constraint:generic} guarantee that the index scales like $N^{5/3}$ in the large $N$ limit \cite{Hosseini:2016tor}.
The Bethe potential $\cV$, whose extremum gives the pole configurations of the contour integral \eqref{index:path integral}, is then given by \cite{Hosseini:2016tor}
\bea
 \label{Bethe:potential:N53}
 \begin{split}
 \frac{\cV \left( \rho(t) , v(t) , \Delta_I \right)}{N^{5/3}} & = n \int {\rm d}t\ \rho(t)\ \left\{ - i t\, v(t) + \frac{1}{2} \left[t^2 - v(t)^2\right] \right\}  \\
 & \phantom{=} + i \sum_{I} g_+ (\Delta_I) \int {\rm d}t \, \frac{\rho(t)^2}{1- i v'(t)} - i \mu \left( \int {\rm d}t\, \rho(t) - 1 \right)\, ,
 \end{split}
\eea
where we introduced the polynomial functions
\bea
 \label{gp:gm}
 g_+(u) = \frac{u^3}6 - \frac\pi2 u^2 + \frac{\pi^2}3 u \, ,
 \qquad\qquad g_{+}'(u) = \frac{u^2}2 - \pi u + \frac{\pi^2}3 \, ,
\eea
assuming that $0 \leq \Delta_I \leq 2 \pi$, and
\bea
 \label{Romans:mass:k}
 n \equiv \sum_{a = 1}^{|G|} k_a = |G| k \, .
\eea
We need to extremize the local functional $\cV\left(\rho(t), v(t) , \Delta_I \right)$
with respect to the continuous functions $\rho(t)$ and $v(t)$. 
The solution for $\sum_{I \in a} \Delta_I = 2 \pi$, for each term $W_a$ in the superpotential,
is as follows:\footnote{The support $[t_-, t_+]$ of $\rho(t)$
can be determined from the relations $\rho(t_\pm)=0$.} 
\bea
 \label{BAEs:sol:generic}
 v(t) & = -\frac{1}{\sqrt{3}} t \, , \\
 \rho(t) & = \frac{3^{1/6}}{2} \left[ \frac{n}{\sum_I g_+(\Delta_I)} \right]^{1/3}
 - \frac{2}{3^{3/2}} \left[ \frac{n}{\sum_I g_+(\Delta_I)} \right] t^2 \, , \\
 t_\pm & = \pm \frac{3^{5/6}}{2} \left[ \frac{\sum_I g_+ (\Delta_I)}{n}\right]^{1/3} \, , \\
 \mu & = \frac{\sqrt{3}}{4} \left( 1 - \frac{i}{\sqrt{3}} \right) n^{1/3}
 \left[ 3 \sum_{I} g_+ (\Delta_I) \right]^{2/3} \, .
\eea
One can explicitly check that
\bea
 \label{BethePot:on-shell}
 \overline\cV (\Delta_I) \equiv - i \cV \Big|_\text{BAEs} = \frac{3}{5} \mu N^{5/3}\, .
\eea
This is indeed equal to $- \log Z_{S^3}$, \cf\;Eq.\,(3.26) in \cite{Fluder:2015eoa}, up to a normalization \cite{Hosseini:2016cyf}.
Here $Z_{S^3}$ is the partition function of the same $\cN=2$ theory on the three-sphere \cite{Kapustin:2009kz,Jafferis:2010un,Hama:2010av}.

For this class of Chern-Simons-matter quiver gauge theories the topologically twisted index, at large $N$, reads \cite{Hosseini:2016tor}
\bea
 \label{index:N53}
 \log Z  = - \left[ |G| \frac{\pi^2}{3} + \sum_{I} (\fn_I - 1) g'_+ (\Delta_I) \right] N^{5/3}
 \int {\rm d}t \, \frac{\rho(t)^2}{1-i v'(t)} \, .
\eea
Plugging the solution \eqref{BAEs:sol:generic} into the index \eqref{index:N53},
we obtain the following simple expression for the logarithm of the index
\bea
 \label{index:generic}
 \log Z \left( \fn_I, \Delta_I \right) = - \frac{3^{7/6}}{10} \left( 1 - \frac{i}{\sqrt{3}} \right) n^{1/3} N^{5/3}
 \frac{\sum_{I} \left[ \frac{3}{\pi} g_{+}(\Delta_I) +
 \left(\fn_I - \frac{\Delta_I}{\pi } \right) g_{+}'(\Delta_I) \right]}{\left[ \sum_{I} g_+(\Delta_I) \right]^{1/3}}
 \, .
\eea
Remarkably, it can be rewritten as
\bea
 \label{index:generic:c2d:a4d}
 \log Z \left( \fn_I, \Delta_I \right) =
 - \frac{ 3^{7/6} \pi}{5 \times 2^{10/3}}
 \left(1 - \frac{i}{\sqrt{3}} \right)
 \left( n N \right)^{1/3}
 \frac{c_r \left( \fn_I , \Delta_I \right)}{a \left( \Delta_I \right)^{1/3}} \, .
\eea
Here $a \left( \Delta_I \right)$ is the trial $a$ central charge of the ``parent'' four-dimensional
$\cN = 1$ superconformal field theory on $S^2 \times T^2$, with a partial topological A-twist on $S^2$,
and $c_r  \left( \fn_I , \Delta_I \right)$ is the trial right-moving central charge of the two-dimensional $\cN = (0,2)$
theory on $T^2$ obtained from the compactification on $S^2$.\footnote{We refer the reader
to \cite{Benini:2012cz,Benini:2013cda,Hosseini:2016cyf,Karndumri:2013dca,Klemm:2016kxw,Amariti:2016mnz,Amariti:2017cyd,Amariti:2017iuz} for a detailed analysis of
superconformal theories obtained by twisted compactifications of four-dimensional $\cN = 1$ theories and their holographic realization.}
They read \cite{Benini:2012cz,Benini:2013cda,Hosseini:2016cyf}
\bea
 \label{a4d:c2d:anomaly}
 a \left( \Delta_I \right) & = \frac{9}{32} \Tr R^3 \left( \Delta_I \right) - \frac{3}{32} \Tr R \left( \Delta_I \right) \, , \\
 c_{r} \left( \fn_I , \Delta_I \right) & =
 3 \Tr R^3 ( \Delta_I) + \pi \sum_{I} \left[ \left( \fn_I - \frac{\Delta_I}{\pi} \right) \frac{\partial \Tr R^3 ( \Delta_I)}{\partial \Delta_I} \right] \, ,
\eea
where the R-symmetry 't Hooft anomaly of four-dimensional $\cN = 1$ SCFTs is given in terms of quiver data
\bea
 \label{tHoof:linear:anomalies}
 \Tr R^{\alpha} (\Delta_I) & =
 |G| \text{ dim } \SU(N) +
 \sum _{I} \text{ dim }\fR_I \left( \frac{\Delta_I}{\pi} - 1 \right)^{\alpha} \, .
\eea
Here $\alpha = 1 , 3$ and $\text{dim }\fR_I$ is the dimension of the respective matter representation with R-charge $\Delta_I / \pi$.
Notice that Eq.\,\eqref{index:generic:c2d:a4d} is consistent with the {\it index theorem} of \cite{Hosseini:2016tor}
\bea
 \label{index:theorem:general}
 \log Z \left( \fn_I, \Delta_I \right) = - \frac{2}{\pi} \overline\cV \left( \Delta_I \right) -
 \sum_{I} \left[ \left( \fn_I - \frac{\Delta_I}{\pi} \right)
 \frac{\partial \overline\cV \left( \Delta_I \right)}{\partial \Delta_I} \right] \, .
\eea

\subsection[The index of D2\texorpdfstring{$_k$}{[k]} at large \texorpdfstring{$N$}{N}]{The index of D2$_{\bm k}$ at large ${\bm N}$}
\label{ssec:SYM-CS:index}

So far the discussion was completely general. Let us now focus on the $\cN = 2$ Chern-Simons
deformation of the maximal SYM theory in three dimensions \cite{Schwarz:2004yj,Guarino:2015jca}.
In $\cN = 2$ notation, the three-dimensional maximal SYM has an adjoint vector multiplet (containing a real scalar and a complex fermion)
with gauge group $\U(N)$ or $\SU(N)$ as well as three chiral multiplets $\phi_j$ $(j = 1, 2, 3)$ (containing a complex scalar and fermion).
This theory has $\U(1)_R \times \SU(3)$ symmetry, with $\SU(3)$ rotating the three complex scalar fields in the chiral multiplets.
The quiver diagram for this theory is depicted below.
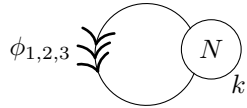
\bea
 \label{SYM:k:quiver}
 \begin{tikzpicture}[font=\footnotesize, scale=0.9]
  \begin{scope}[auto,%
   every node/.style={draw, minimum size=0.5cm}, node distance=2cm];
  \node[circle]  (UN)  at (0.3,1.7) {$N$};
  \end{scope}
  \draw[decoration={markings, mark=at position 0.45 with {\arrow[scale=2.5]{>}}, mark=at position 0.5 with {\arrow[scale=2.5]{>}}, mark=at position 0.55 with {\arrow[scale=2.5]{>}}}, postaction={decorate}, shorten >=0.7pt] (-0,2) arc (30:341:0.75cm);
  \node at (-2.2,1.7) {$\phi_{1,2,3}$};
  \node at (0.7,1.2) {$k$};
 \end{tikzpicture}
 \nonumber
\eea
It has a cubic superpotential,
\bea
 W = \Tr \left( \phi_3 \left[ \phi_1 , \phi_2 \right] \right) \, .
\eea
We assign chemical potentials $\Delta_j \in [0 , 2 \pi]$ to the fields $\phi_{j}$.
The invariance of each monomial term in the superpotential under the global symmetries of the theory imposes the following constraints
on the chemical potentials $\Delta_j$ and the flavor magnetic fluxes $\fn_j$ associated with the fields $\phi_j$,
\bea
 \sum_{j = 1}^{3} \Delta_j \in 2 \pi \bZ \, , \qquad \qquad \qquad
 \sum_{j = 1}^{3} \fn_j = 2 \, ,
\eea
where the latter comes from supersymmetry.
Since $0 \leq \Delta_j \leq 2\pi$ we can only have $\sum_{j=1}^{3} \Delta_j=2 \pi s\, ,\forall s=0,1,2,3$.
The cases $s=0,3$ are singular while those for $s=2$ and $s=1$ are related by a discrete symmetry $\Delta_j = 2\pi - \Delta_j$.
Thus, without loss of generality, we will assume $\sum_{j = 1}^{3} \Delta_j = 2 \pi$.
We find that
\bea
 \sum _{j = 1}^3 g_+ \left( \Delta_j \right) & = \frac{1}{2} \Delta_1 \Delta_2 \Delta_3 \, , \\
 \sum _{j = 1}^3 g_+' \left( \Delta_j \right) & =
 \frac{1}{4} \left[ \left( \Delta_1^2 + \Delta_2^2 + \Delta_3^2 \right)
 - 2 \left(\Delta_1 \Delta_2 + \Delta_2 \Delta_3 + \Delta_1 \Delta_3 \right) \right] \, .
\eea
Finally, the ``on-shell'' value of the Bethe potential \eqref{BethePot:on-shell} and the index \eqref{index:generic}, at large $N$, can be written as
\be
 \label{index:SYM-CS}
 \begin{aligned}
 \overline\cV \left( \Delta_j \right) & =
 \frac{3^{13/6}}{5 \times 2^{8/3}}
 \left(1 - \frac{i}{\sqrt{3}} \right)
 k^{1/3} N^{5/3}
 \left( \Delta_1 \Delta_2 \Delta_3 \right)^{2/3} \, , \\
 \log Z \left( \fn_j , \Delta_j \right) & = - \frac{3^{7/6}}{5 \times 2^{5/3}}
 \left(1 - \frac{i}{\sqrt{3}} \right)
 k^{1/3} N^{5/3}
 \left( \Delta_1 \Delta_2 \Delta_3 \right)^{2/3}
 \sum_{j=1}^3 \frac{\fn_j}{\Delta_j}
 \, ,
 \end{aligned}
\ee
which is valid for $\sum_{j=1}^{3} \Delta_j = 2 \pi$ and $0 \leq \Delta_j \leq 2 \pi$.
Note also that
\bea
 \label{index:SYM:k:attractor}
 \log Z \left( \fn_j , \Delta_j \right) = - \sum_{j = 1}^{3} \fn_j \frac{\partial \overline{\cV}\left(\Delta_j\right)}{\partial \Delta_j} 
 \, ,
\eea
as expected from the index theorem \eqref{index:theorem:general}.

\section[AdS\texorpdfstring{$_{\bm{4}}$}{(4)} black holes in \texorpdfstring{$\bm{\mathcal{N}=2}$}{N=2} dyonic STU  gauged supergravity]{AdS$_{\bm 4}$ black holes in $\bm{\mathcal{N}=2}$ dyonic STU  gauged supergravity}
\label{sec:dyonic sugra}

We now turn to the gravity duals of the field theories we have discussed so far. Following the logic and analysis of \cite{Benini:2015eyy}, where a successful microscopic entropy counting for AdS$_4 \times S^7$ black holes was performed at leading order in the large $N$ limit, we search for four-dimensional black hole solutions which interpolate between an AdS$_2 \times \Sigma_{\mathfrak{g}}$ near-horizon region and an asymptotic AdS$_4$ vacuum. These preserve supersymmetry due to a topological twist on the Riemann surface $\Sigma_{\mathfrak {g}}$. We find that the gravity dual is a black hole solution which can be embedded in massive type IIA supergravity and is asymptotic to the $\mathcal{N} = 2$ supersymmetric AdS$_4 \times S^6$ background of \cite{Guarino:2015jca}. 

The instance of AdS$_4$/CFT$_3$ correspondence we aim to explore here was finalized in \cite{Guarino:2015jca} and states that the D2$_k$ theory admits a dual description as massive type IIA string theory on asymptotically AdS$_4 \times S^6$ backgrounds. Of particular importance is the consistent truncation of massive type IIA supergravity on $S^6$, to the dyonically gauged maximal supergravity in four dimensions with ISO$(7)$ gauge group \cite{Guarino:2015jca, Guarino:2015vca}.

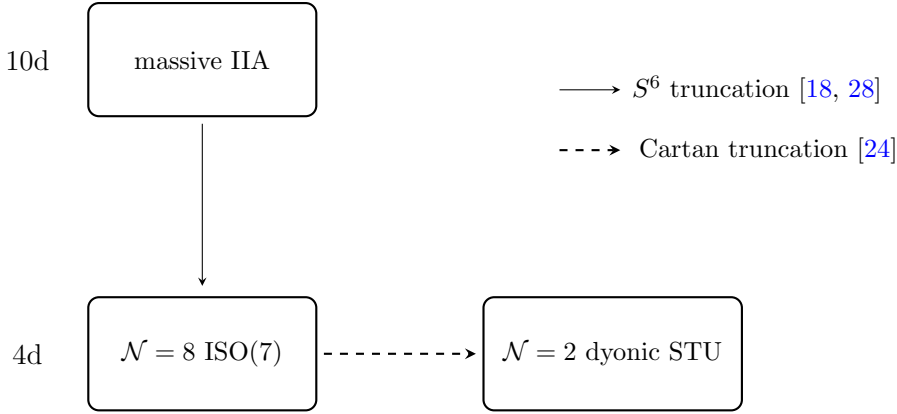
\begin{figure}[ht]
\centering	
\begin{tikzpicture}[scale=1, every node/.style={scale=.9}]
	
	\node at (-2.3,0){10d};
	
	\draw[fill = white,thick, rounded corners] (-1.5,-.75) rectangle (1.5,.75);
	\node at (0,0){\small massive IIA};
	
	\draw[->,>=stealth] (0,-.85) -- (0,-3);
		
	\node at (-2.3,-3.9){4d};
		
	\draw[fill = white,thick, rounded corners] (-1.5,-4.65) rectangle (1.5,-3.15);
	\node at (0,-3.9){\small $\mathcal{N}=8$ ISO(7)};
		
	\draw[->,>=stealth, thick, dashed] (1.6,-3.9) -- (3.6,-3.9);	
				
	\draw[fill = white,thick, rounded corners] (3.7,-4.65) rectangle (7.1,-3.15);
	\node at (5.4,-3.9){\small $\mathcal{N}=2$ dyonic STU };

	\draw[->,>=stealth, ]  (4.7, -.4) -- (5.5, -.4);
	\node at (7.3,-.4){\small $S^6$ truncation \cite{Guarino:2015jca,Guarino:2015vca}};
	
	\draw[->,>=stealth, ,dashed, thick] (4.7,-1.2) -- (5.5,-1.2);
	\node at (7.45,-1.2){\small Cartan truncation \cite{Guarino:2017pkw}};
		
	\end{tikzpicture}
	\caption{Sequence of consistent truncations from massive type IIA supergravity in ten dimensions, to the dyonic STU model in four dimensions.}
	\label{trunc_fig}
\end{figure}

Before discussing the main features of the supergravity theory we consider, let us elaborate on the desired type of black hole solutions. The first examples of such analytic solutions appeared in \cite{Cacciatori:2009iz} and were further studied in \cite{DallAgata:2010ejj, Hristov:2010ri,Halmagyi:2013sla,Katmadas:2014faa,Halmagyi:2014qza}. They are solutions of ${\cal N} = 2$ supergravity with a gauged U$(1)$ R-symmetry group. The aforementioned topological twist consists of the cancellation of the spin connection on $\Sigma_{\mathfrak g}$ by the R-symmetry gauge vector field, and requires that the black hole solutions carry non-trivial abelian magnetic charges. 
In the case of the maximally supersymmetric SO$(8)$ gauged supergravity this allows one to consider only an abelian $\U(1)^4$ truncation (often called ``STU model''), as was the case in \cite{Benini:2015eyy}.

These considerations lead us to look at the $\mathcal{N} = 2$ truncation of the dyonic ISO$(7)$-gauged supergravity constructed recently in \cite{Guarino:2017pkw}, as the analogue of the STU model. It corresponds to picking the maximal abelian subgroup of the original ISO$(7)$ gauge group and arranging the resulting four abelian vectors in an ${\cal N}=2$ gravity multiplet and three vector multiplets. Due to the characteristics of the supergravity theory under consideration, there is the requirement that the four vectors couple to a hypermultiplet, so that they effectively gauge some of the isometries of the scalar manifold. The details can be found in the original reference \cite{Guarino:2017pkw}. Here we review the main data for constructing uniquely the supergravity Lagrangian; more details can be found in appendix \ref{appendix}. The three complex scalars in the three vector multiplets parameterize the special K\"{a}hler scalar (SK) manifold ${\cal M}_{{\rm SK}} = [{\rm SU}(1,1)/{\rm U}(1)]^3$ and the additional hypermultiplet, the so-called universal hypermultiplet (often arising in string theory compactifications), parameterizes the quaternionic K\"{a}hler (QK) manifold ${\cal M}_{{\rm QK}} = {\rm SU}(2,1)/({\rm SU}(2)\times {\rm U}(1))$.

We follow the conventions of \cite{Andrianopoli:1996cm} adapted to the most general ${\cal N}=2$ dyonically gauged supergravity theory spelled out in \cite{deWit:2011gk}: the vector multiplet sector is defined by a prepotential
\begin{equation}\label{prepot}
	{\cal F} = - 2 \sqrt{X^0 X^1 X^2 X^3}\ ,
\end{equation}
where the holomorphic sections $X^{\Lambda}$, $\Lambda = 0,1,2,3$ parameterize the SK manifold. The symplectic vector $ (X^{\Lambda}, F_{\Lambda} \equiv \partial {\cal F}/\partial X^{\Lambda})$ can be parameterized by three complex scalars $s, t$ and $u$ as
\begin{align}\label{parameterization}
	(X^0,\, X^1,\, X^2,\, X^3,\, F_0,\, F_1,\, F_2,\, F_3) = (- s t u,\, -s,\, -t,\, -u,\, 1,\, t u,\, s u,\, s t)\ .  
\end{align}
The QK manifold is parameterized by four real scalars $\{ \phi,\, \sigma,\, \zeta,\, \tilde{\zeta} \}$.
The four vector fields are used to gauge two commuting isometries of the universal hypermultiplet metric,\footnote{See, for example, appendix D of \cite{Hristov:2012bk} for an explicit realization of all eight isometries of the universal hypermultiplet metric forming the group SU$(2,1)$.} generated by the Killing vectors
\begin{align}
k_0 = \partial_\sigma \ , \qquad 
k^0 = c \partial_\sigma \ , \qquad 
k_{1,2,3} = - \tilde{\zeta} \partial_\zeta + \zeta \partial_{\tilde{\zeta}} \ , \qquad
k^{1,2,3} = 0 \ ,
\end{align}
where the indices $0$ to $3$ refer to the four vectors; a lower index is associated to the ordinary (electric) gauge field, whereas an upper index to the dual (magnetic) gauge field, as is the standard to keep track of symplectic covariance. The constants $m$ and $g$ correspond to the magnetic and electric gauge coupling constants respectively, and in what follows we will make use of the ratio
\begin{equation}\label{c}
	c \equiv \frac{m}{g}\ .
\end{equation}
Note that although all four abelian vectors participate in the gauging of the above isometries, only two different isometries are actually being gauged: one corresponding to the non-compact group $\mathbb{R}$, gauged by a linear combination of the electric and magnetic gauge fields $A^0$ and $A_0$, and a U$(1)$ isometry gauged by the linear combination $A^1 + A^2 + A^3$. These gaugings, via supersymmetry, generate a non-trivial scalar potential, which has a critical point corresponding to an AdS$_4$ vacuum. 

Our aim is therefore to find supersymmetric black hole solutions in the theory we just described. We will do so in several steps, leaving all detailed calculations to the appendix; for each step we find useful to dedicate a subsection. First, we describe the black hole ansatz and supersymmetry equations derived by \cite{Halmagyi:2013sla,Klemm:2016wng}. We then concentrate separately on the conditions for the asymptotic AdS$_4$ vacuum and the near-horizon AdS$_2 \times \Sg$ geometry. We manage to rewrite the near-horizon data in a particularly simple form in order to facilitate the match with the field theory. We finish the supergravity analysis by presenting arguments for the existence of a full BPS flow between the UV and IR geometry. Ultimately, the existence of the complete geometries is best justified by the successful entropy match with field theory.

\subsection{Black hole ansatz and BPS conditions}
Static BPS AdS$_4$ black holes in general models with dyonic hypermultiplet gauging,  were considered in  \cite{Klemm:2016wng} generalizing earlier work of \cite{Cacciatori:2009iz,DallAgata:2010ejj,Hristov:2010ri,Halmagyi:2013sla,Katmadas:2014faa,Halmagyi:2014qza}. The reader can find all the details about the bosonic ansatz and BPS equations in appendix \ref{appendix}. Here, for the sake of clarity, we repeat the form of the metric,
\begin{equation}
{\rm d} s^2 = - \e^{2U(r)} {\rm d} t^2 + \e^{-2U(r)} {\rm d} r^2 + \e^{2(\psi(r) - U(r))}{\rm d}\Omega_{\kappa}^2\,,
\end{equation}
where the radial functions $U(r)$, $\psi(r)$ and the choice of scalar curvature $\kappa$ for the horizon manifold, uniquely specify the spacetime. Electric and magnetic charges, $e_{\Lambda} (r)$ and $p^{\Lambda} (r)$, are present for each gauge field, and can have a radial dependence due to the fact that some of the hypermultiplet scalars source the Maxwell equations. The spacetime symmetries also impose a purely radial dependence for the SK complex scalars $s (r)$, $t (r)$, $u (r)$ and the QK real scalars $\phi (r)$, $\sigma (r)$, $\zeta (r)$, $\tilde{\zeta} (r)$, as well as the phase $\alpha (r)$ of the Killing spinors that parameterize the fermionic symmetries of the black hole. 

We systematically write down the conditions for supersymmetry and equations of motion in the appendix, while here we only discuss the most important points about the solution. In particular we find that we can already fix three of the four hypermultiplet scalars
\begin{equation}
	\zeta = \tilde{\zeta} = 0\ , \qquad \sigma = \text{const.} \, ,
\end{equation}
where the particular value of $\sigma$ is not physical as it is a gauge dependent quantity that drops out of all BPS equations. The remaining hypermultiplet scalar however has in general a non-trivial radial profile governed by the equation
\begin{equation}\label{phi}
	\phi' = - g \kappa \lambda \e^{{\cal K}/2 - U} {\rm Im} \left(\e^{-i \alpha} (X^0 - c F_0)\right)\ , 	
\end{equation}
where the K\"ahler potential $\e^{\cal K}$ and $\lambda = \pm 1$ are discussed in the appendix.
The scalar $\phi$ precisely sources the Maxwell equations, which read 
\begin{equation}\label{charge}
	p'^0 = c e'_0 = -c \e^{2 \psi - 3 U} \e^{4 \phi} {\rm Re}  \left(\e^{-i \alpha} (X^0 - c F_0)\right)\ ,
\end{equation}
while all other charges $p^{1,2,3}$ and $e_{1,2,3}$ are truly conserved quantities. These two equations highlight an important physical feature of the black holes in massive IIA supergravity: due to the presence of charged hypermultiplet scalars there are massive vector fields that do not have conserved charges. The charges of the massive vectors are not felt by the field theory, which explains why there were only three different charges considered in the previous section. These are the magnetic charges $p^{1,2,3}$ as here we will further simplify our ansatz and put $e_{1,2,3}= e$ to be fixed by the magnetic charges. However, one still needs to solve consistently the BPS equations for the massive vector fields, which presents a particularly hard obstacle computationally, and has prevented people from writing down exact analytic solutions for black holes with massive vector fields before \cite{Halmagyi:2013sla}.

\subsection{Constant scalars, analytic UV and IR geometries}
Let us now concentrate on the two important end-points of the full black hole flow: the asymptotic UV space AdS$_4$ and the IR fixed point, AdS$_2 \times \Sg$. Due to the symmetries of these spaces the scalars and charges are constant there, which means \eqref{phi}-\eqref{charge} can be further constrained by setting their left-hand sides to zero. This immediately implies 
\begin{equation}
	X^0 - c F_0 = 0 \Rightarrow X^0 = (- c)^{2/3} (X^1 X^2 X^3)^{1/3}\ , \qquad stu = -c\ , 
\end{equation}
which presents a strong constraint of the vector multiplet moduli space. In fact the remaining scalars (\eg\;freezing $s$ in favor of $t$ and $u$) are consistent with the simplified prepotential\footnote{Note that directly substituting $X^0$ in the original prepotential \eqref{prepot} leads to a different normalization. Such a different prefactor does not lead to a change in physical quantities, but we prefer to comply with the correct normalization of the kinetic terms as imposed by the choice of parameterization in \eqref{parameterization}.}
\begin{equation}\label{new_prepot}
	{\cal F}^\star = - \frac{3}{2} (-c)^{1/3} (X^1 X^2 X^3)^{2/3}\ .
\end{equation}

Notice that in this constant scalar case, the BPS equations automatically lead us to an effective truncation of the theory to a subsector, by ``freezing'' some of the fields. In particular, we see that the massive vector field has ``eaten up'' the Goldstone boson $\sigma$, and together with the massive scalars $\zeta$, $\tilde{\zeta}$, $\phi$ and the complex combination of $s t u$ can be integrated out of the model. This corresponds to a supersymmetry preserving version of the Higgs mechanism discussed in \cite{Hristov:2010eu} and a truncation\footnote{Note that strictly speaking we have not proven that this is a consistent truncation as the proof in \cite{Hristov:2010eu} only considered electrically gauged hypermultiplets. For the analogous proof in the general dyonic case one needs to use the full superconformal formalism of \cite{deWit:2011gk} where the general theory is properly defined. However, here we never need to go to such lengths since we use the Higgs mechanism to clarify the physical picture, not as a guiding principle in deriving the BPS equations.} to an ${\cal N}=2$ theory with two massless vector multiplets and no hypermultiplets. The remainders of the gauged hypermultiplet are constant parameters gauging the R-symmetry, known as Fayet-Iliopoulos terms, $\xi_I = P^{x=3}_I$, $I \in \{1,2,3 \}$. Therefore the effective, or truncated, prepotential ${\cal F}^\star$ is indeed the prepotential defining the Higgsed theory. This mechanism is in fact the reason why we are able to write down exact analytic solutions in the UV and IR limits where the constant scalar assumptions holds. Note that one could have in principle performed this truncation of the full theory looking for full black hole solutions there. However, this turns out to be a too strong constraint; in particular we will see that in the UV we have 
\begin{equation}
	\langle \e^{2 \phi} \rangle_{\rm UV} = 2 c^{-2/3}\ ,
\end{equation}  
while in the IR in general
\begin{equation}
	\langle \e^{2 \phi}\rangle_{\rm IR} = \frac{2 c^{-2/3}}{3 (H^1 H^2 H^3)^{1/3}}\ ,
\end{equation}  
with $H^I$ particular functions of the charges. Imposing the constraint that $\phi$ is constant throughout the flow $\phi_{\rm UV} = \phi_{\rm IR}$ leads to a black hole solution with only a subset of all possible charges, in particular to the so called universal twist dating back to \cite{Caldarelli:1998hg} and studied for this theory in \cite{Guarino:2017eag}, recently described holographically by \cite{Azzurli:2017kxo}.

\subsubsection[Asymptotic AdS\texorpdfstring{$_4$}{(4)} vacuum]{Asymptotic AdS$_4$ vacuum}
As already noticed in \cite{Hristov:2011ye} the black hole is only asymptotically {\it magnetic} AdS$_4$, which means that the black hole solution even asymptotically never recovers the full AdS symmetry. For this reason the BPS equations are only solved at $r \rightarrow \infty$. We can nevertheless analyze the exact AdS$_4$ vacuum, which constrains the scalar fields to obey the maximally supersymmetric conditions derived in \cite{Hristov:2010eu}. These conditions, as shown in more detail in the appendix, not only constrain the scalars to be constant with $\zeta = \tilde{\zeta} = 0$ and $s t u= -c$ but further impose the particular vacuum expectation values
\begin{align}\label{vacuum}
\begin{split}
\langle s\rangle _{{\rm AdS}_4} = \langle t\rangle _{{\rm AdS}_4} = \langle u\rangle _{{\rm AdS}_4} = (-c)^{1/3},\\
\langle \e^{2 \phi}\rangle _{{\rm AdS}_4} = 2 c^{-2/3}\ ,
\end{split}
\end{align} 
which can be checked to explicitly solve all the equations \eqref{app:BHequations} at $r \rightarrow \infty$. The metric functions in this limit become
\begin{equation}
\label{AdS4:radius}
	\lim_{r \rightarrow \infty} (r \e^{-\psi}) = \lim_{r \rightarrow \infty} \e^{-U} = \frac{L_{{\rm AdS}_4}}{r}\ , \quad L_{{\rm AdS}_4} = \frac{c^{1/6}}{3^{1/4} g}\ , 
\end{equation}
as already found in \cite{Guarino:2017eag}.

\subsubsection{Near-horizon geometry and attractor mechanism}
The attractor mechanism for static supersymmetric asymptotically AdS$_4$ black holes was studied in detail in \cite{Klemm:2016wng}, generalizing the results of \cite{DallAgata:2010ejj} to cases with general hypermultiplet gaugings.
The near-horizon geometry is of the direct product type AdS$_2 \times \Sg$ and preserves four real supercharges, double the amount preserved by the full black hole geometry. We solve carefully all equations in the appendix, while here we present an alternative derivation which, although incomplete as we explain in due course, is more suitable for the comparison with field theory.

The near-horizon metric functions are given by
\begin{equation}
U = \log(r/L_{{\rm AdS}_2}) \, , \qquad \psi = \log(L_{\Sigma_\fg} \cdot r / L_{{\rm AdS}_2}) \, ,
\end{equation}
where $L_{{\rm AdS}_2}$ is the radius of AdS$_2$ and $L_{\Sigma_\fg}$ that of the surface $\Sigma_\fg$.

We start with the BPS condition coming from the topological twist for the magnetic charges (valid not only on the horizon but everywhere in spacetime)
\begin{equation}
 g \sum_{I=1}^3 p^I = - \kappa\ ,
\end{equation}
with $\kappa$ the unit curvature of the internal manifold on the horizon ($\kappa = +1$ for $S^2$ and $\kappa = -1$ for $\Sigma_{\mathfrak{g}>1}$). The general attractor equations imply in particular that the horizon radius is given by
\begin{align}\label{entropy extremization}
	L^2_{\Sg} =  i \kappa \frac{\cal Z}{\cal L} = - i \frac{ \sum_I ( e_I X^I - p^I F_I) }{g (X^1+X^2+X^3)}\ . 
\end{align}
where in the last equality we already used the model specific information that $X^0 - c F_0 = 0$ which implies $ X^0 = (- c)^{2/3} (X^1 X^2 X^3)^{1/3}$. Notice that the same equation is found by directly using the truncated prepotential, ${\cal F}^\star$, since by construction  
\begin{align}
\begin{split}
F_I ( X^0 &= (- c)^{2/3} (X^1 X^2 X^3)^{1/3}) = F^\star_I\ , \quad \forall I \in \{1,2,3\}\ , \\
\Rightarrow L^2_{\Sg} &=  i \kappa \frac{\cal Z^\star}{\cal L^\star}\ .
\end{split}
\end{align}
This shows that we can equally well use the truncated prepotential for this attractor equation. To solve it, we define the weighted sections $\hat{X}^I \equiv X^I/\sum_J X^J$ such that $\sum_I \hat{X}^I = 1$, and find
\begin{align}
\label{sugra:index}
	{\sum_{I=1}^{3}} \left( p^I \hat{F}^\star_I  -  e_I \hat{X}^I \right) = g L^2_{\Sg}\ , 
\end{align}
where we used the shorthand notation $F_I^\star (\hat{X}^I) \equiv \hat{F}^\star_I$. This expression is extremized at the horizon
\begin{align}\label{sugra:extremization}
	\left. \partial_{\hat{X}^J}\left[ \tsumI (p^I \hat{F}^\star_I - e_I \hat{X}^I) \right] \right|_{\hat{X}_{\rm horizon}} = 0\ , 
\end{align}
fixing the weighted sections, $\hat{X}^I_{\rm horizon} \equiv H^I$ in terms of the electric and magnetic charges. 

Let us now concentrate on what we call ``purely magnetic'' solution, \ie\;let us work under the assumption that we only have independent magnetic charges and all electric charges are equal $e_I = e$. The equations simplify to
\begin{align}\label{extremization}
	\left. \partial_{\hat{X}^J} \left( \tsumI p^I \hat{F}^\star_I \right)\right|_{H^I}  = 0\ , 
\end{align}
given $\sum_I \hat{X}^I = 1$. We find the following solutions:
\begin{equation}\label{Xsolution}
 3 H^I = 1 \pm \sum_{J, K} \frac{\left| \epsilon_{I J K} \big( p^J - p^K \big) \right|}{2 \sqrt{\big( \sqrt{\Theta} \pm p^I \big)^2 - p^J p^K}}\, ,
\end{equation}
where the $\pm$ signs are not correlated so we have four solutions.
Here $\epsilon_{I J K}$ is the Levi--Civita symbol and  
\begin{equation}\label{theta}
\Theta (p) \equiv \left(p^1\right)^2+\left(p^2\right)^2+\left(p^3\right)^2 - \left( p^1 p^2 + p^1 p^3 + p^2 p^3 \right) .
\end{equation}
The sign ambiguities are to be resolved in the full geometry as proper normalization of the scalar kinetic terms require that $0 < {\rm Im}(s, t, u)<1$ everywhere in spacetime, including the horizon values. It is now straightforward to derive the physical scalars from the weighted sections $H^I$,
\begin{equation}
	s = \frac{\e^{i \pi/3} c^{1/3} H^1}{(H^1 H^2 H^3)^{1/3}}\ , \quad t = \frac{\e^{i \pi/3} c^{1/3} H^2}{(H^1 H^2 H^3)^{1/3}}\ , \quad u = \frac{\e^{i \pi/3} c^{1/3} H^3}{(H^1 H^2 H^3)^{1/3}}\ .
\end{equation}
At first it might seem that there is an ambiguity in the attractor equation, since at the moment we have allowed for an arbitrary parameter $e$ which sets the value of the three equal electric charges. This is however misleading, because we have in fact not yet solved the original equation \eqref{entropy extremization}. The electric charges there play the crucial r\^ole of making sure the radius of the horizon is indeed a positive real quantity,
\begin{align}
\begin{split}
\frac{\cal Z}{\cal L} &= -\frac{\kappa}{g} \left( (-1)^{4/3} c^{1/3} (H^1 H^2 H^3)^{2/3} \tsumI(p^I/H^I) - \tsumI e_I H^I \right) \\
&= -\frac{\kappa}{g} \left(\e^{-2 i \pi/3} c^{1/3} (H^1 H^2 H^3)^{2/3} \tsumI  (p^I/H^I )  - e \right)  \\
&= - i \kappa L^2_{\Sg}\ .
\end{split}
\end{align}
The imaginary part of the last equation fixes the radius of the Riemann surface,
\begin{equation}\label{area}
	L^2_{\Sg} = -\frac{\sqrt{3}}{2 g} c^{1/3} (H^1 H^2 H^3)^{2/3} \sum_{I=1}^3 \frac{p^I}{H^I} \, 
\end{equation}
while the real part fixes the value of the electric charges,
\begin{equation}
	e = \frac{1}{2} c^{1/3} (H^1 H^2 H^3)^{2/3} \sum_{I=1}^3 \frac{p^I}{H^I}  = - \frac{g}{\sqrt{3}} L^2_{\Sg} \ .
\end{equation}
However, \eqref{entropy extremization} can only get us this far, and one needs to solve the other near-horizon equations in order to write down the full solutions, as we have done in the appendix. This way one can fix the massive vector charges $p^0$, $e_0$, as well as the hypermultiplet scalar $\phi$:
\begin{equation}
	p^0 = c e_0 = \frac{g c^{1/3}}{3 \sqrt{3} (H^1 H^2 H^3)^{1/3}} L^2_{\Sg}\ ,  \qquad \e^{2 \phi} = \frac{2 c^{-2/3}}{3 (H^1 H^2 H^3)^{1/3}}\ .
\end{equation}
The AdS$_2$ radius is also fixed from the remaining near-horizon BPS equations analyzed in the appendix, and it can also be expressed in terms of the functions $H^I$ as
\begin{equation}
	L_{{\rm AdS}_2} = \frac{3^{3/4} c^{1/6} (H^1 H^2 H^3)^{1/3}}{2 g}\ .
\end{equation}
Finally, for completeness, we write the Bekenstein-Hawking entropy for black holes with spherical horiozn $(\kappa=+1)$:\footnote{A precise
counting of microstates for $\fg=0$ case implies matching of the index and the entropy for all values of $\fg$ (see section 6 of \cite{Benini:2016hjo}).}
\be
 \label{BH:entropy:final}
 S_{\rm BH} = \frac{\text{Area}}{4 G_{\rm N}} = \frac{\pi L_{S^2}^2}{G_{\rm N}}
 = - \frac{\pi \sqrt{3}}{2 g G_{\rm N}} c^{1/3} (H^1 H^2 H^3)^{2/3} \sum_{I=1}^{3} \frac{p^I}{H^I} \, .
\ee

\subsection{Existence of full black hole flows}
The main challenge in constructing the full black hole spacetime interpolating between the UV and IR geometries we presented above, is the non-trivial massive vector field we need to consider. We have seen that in the constant scalar case we can effectively decouple the massive vector multiplet but this is not the case for the full flow, if we wish to have the most general spacetime. For the BPS equations, it is useful to define the function
\begin{equation}
	\gamma (r) \equiv c F_0 - X^0 = c + s(r) t(r) u(r)\ ,
\end{equation} 
which vanishes both in the UV and the IR. The function $\gamma (r)$ is in principle fixed by the BPS equations determining the scalars $s$, $t$, and $u$, and in turn governs the flow of the hypermultiplet scalar field $\phi$ as well as the massive vector charge $p^0$ via \eqref{phi} and \eqref{charge}, respectively. The remaining first order BPS equations involve also the metric functions $U$ and $\psi$, as well as the Killing spinor phase $\alpha$ while the conserved charges $e_{1,2,3}$ and $p^{1,2,3}$ remain constant and have been fixed already at the horizon. Therefore we have a total of eight coupled differential equations for eight independent variables\footnote{Note that in the ``purely magnetic'' ansatz the phase of the complex scalars has been fixed, therefore we count $s$, $t$, $u$ as each is carrying a single degree of freedom.} $\{ s$, $t$, $u$, $\phi$, $p^0$, $U$, $\psi$, $\alpha \}$. All these fields have been uniquely fixed in the UV and IR as shown above and more carefully in the appendix. A similar set of equations with running hypermultiplet scalars has been considered in \cite{Halmagyi:2013sla} with the result that one can always connect the UV and IR solutions with a full numerical flow, whenever the number of free parameters matches the number of first order differential equations, as is also the case here. It is of course interesting to find such solutions explicitly but we leave this for a future investigation as the main scope here is the field theory match of our results, to which we turn now.

\section{Comparison of index and entropy}
\label{sec:index vs entropy}

In this section we shall derive the Bekenstein-Hawking entropy \eqref{BH:entropy:final} --- including the numerical factor ---
for a class of dyonic BPS black holes discussed in the previous section
from a microscopic counting of supersymmetric ground states in ${\rm D2}_k$, at the leading order $N^{5/3}$.

\subsection[The \texorpdfstring{$\cI$}{I}-extremization principle]{The $\bm{\cI}$-extremization principle}

The Bekenstein-Hawking entropy of a dyonic BPS black hole in AdS$_4$ with a charge vector $(\fn_j, q_j)$
can be obtained by extremizing $\cI ( \Delta_j ) \equiv \log Z \left( \Delta_j \right) - i \sum_{j=1}^{3} \Delta_j q_j$, in the large $N$ limit,
\bea
 \label{SCFT:extremization}
 \frac{\partial \cI \left( \Delta_j \right) }{\partial \Delta_{2,3}} \bigg|_{\sum_{j} \Delta_j = 2 \pi} (\bar \Delta_j) = 0 \, ,
\eea
and evaluating it at its extremum $\bar \Delta_j$:
\bea
 \label{I-extremization:Legendre}
 \cI \big|_{\text{crit}} ( \fn_j , q_j ) =
 S_{\rm BH} ( \fn_j , q_j) \, ,
\eea
with a constraint on the charges that the entropy be real positive.
This procedure, dubbed $\cI$-\emph{extremization} in \cite{Benini:2015eyy,Benini:2016rke}, comprises two steps:
\begin{enumerate}
 \item Extremizing the index unambiguously determines the exact R-symmetry in the unitary $\cN=2$ superconformal quantum mechanics in the IR.
 \item The value of the index at its extremum is the regularized number of ground states.
\end{enumerate}

Interestingly, this is equivalent to the attractor mechanism for BPS black holes, see \eqref{sugra:index}-\eqref{sugra:extremization},
in $\cN = 2$ gauged supergravity in four dimensions \cite{Cacciatori:2009iz,DallAgata:2010ejj,Hristov:2010ri,Klemm:2016wng}.\footnote{See also \cite{Hosseini:2017mds}
for an extremization principle for the entropy of supersymmetric AdS$_5$ black holes.}

\subsection[AdS\texorpdfstring{$_4 \times S^6/$}{(4) x S**6/}D2\texorpdfstring{$_k$}{[k]} correspondence]{AdS$_{\bm 4} \times \bm{S^6/}$D2$_{\bm k}$ correspondence}

Now we are in a position to confront the topologically twisted index of ${\rm D2}_k$,
to leading order in $N$, \eqref{index:SYM-CS} with the Bekenstein-Hawking entropy \eqref{BH:entropy:final}. Let us first note that the relations between SCFT parameters $(N,k)$ and their supergravity duals in massive type IIA,
to leading order in the large $N$ limit, read\footnote{See for example \cite{Fluder:2015eoa,Guarino:2015jca}.}
\bea
 \label{free energy}
 \frac{1}{2 G_{\rm N}} = \frac{2^{1/3} 3^{1/6}}{5} k^{1/3} N^{5/3} \, , \qquad \qquad
 \frac{
 m
 }
 {g}
 = \left( \frac{3}{16 \pi^{3}} \right)^{1/5} k N^{1/5} \, .
\eea
We may choose the magnetic coupling constant
\bea
 \label{electric:coupling}
 m = 3^{3/2} g^7 \, ,
\eea
such that the UV AdS$_4$ metric has unit radius of curvature (see Eq.\;\eqref{AdS4:radius}).

From here on we set $q_j = q\, ,\forall j=1,2,3$.
The topologically twisted index of ${\rm D2}_k$ \eqref{index:SYM-CS} as a function of $\Delta_{2,3}$ is extremized for
\bea
 \label{hatDelta}
 \frac{3 \bar\Delta_{2}}{2 \pi} = 1 \mp \frac{\left|\fn_3 - \fn_1\right|}{\sqrt{\big( \sqrt{\Theta} \pm \fn_2 \big)^2 - \fn_1 \fn_3}} \, , \qquad \qquad
 \frac{3 \bar\Delta_{3}}{2 \pi} = 1 \mp \frac{\left|\fn_1 - \fn_2\right|}{\sqrt{\big( \sqrt{\Theta} \pm \fn_3 \big)^2 - \fn_1 \fn_2}} \, ,
\eea
where we defined the quantity
\bea
 \Theta \equiv \fn_1^2 + \fn_2^2 + \fn_3^2 - \left( \fn_1 \fn_2 + \fn_1 \fn_3 + \fn_2 \fn_3 \right) \, ,
\eea
which is symmetric under permutations of $\fn_j$. Upon identifying
\bea
 \label{Delta:Sections}
 \frac{\bar\Delta_j}{2 \pi} & = H^j \, , \\
 \fn_j & = 2 g p^j \, , \qquad q_j = - \frac{e_j}{2 g G_{\rm N}} \, , \quad \text{ for } \quad j=1,2,3 \, ,
\eea
\eqref{hatDelta} are precisely the values of the weighted holomorphic sections $\hat{X}^j$ at the horizon \eqref{Xsolution}.
The constraint $\sum_{j} \Delta_j \in 2 \pi \bZ$ is consistent with $\sum_j \hat{X}^j = 1$ valid in the bulk.
Plugging the values for the critical points \eqref{hatDelta} back into the Legendre transform of the partition function \eqref{index:SYM-CS},
and employing \eqref{free energy} and \eqref{electric:coupling} we finally arrive at the conclusion
that \eqref{I-extremization:Legendre} holds true.
We thus found a precise statistical mechanical interpretation of the black hole entropy \eqref{BH:entropy:final}.
Obviously, the above discussion goes through for the most general case with three unequal electric charges
and different horizon topologies.

It is worth stressing that the imaginary part of the partition function \eqref{index:SYM-CS} uniquely fixes the value of the electric charges $q_j = q\, , \forall j=1,2,3$ such that its value at the critical point is a real positive quantity in agreement with the supergravity attractor mechanism and the general expectations in \cite{Benini:2016rke}. This precise holographic match therefore presents a new and successful check on the $\cI$-extremization principle in the presence of a non-trivial phase which is new with respect to previous examples such as the index of ABJM.

\section*{Acknowledgements}

We would like to thank Francesco Azzurli, Nikolay Bobev, Giuseppe Dibitetto, Dietmar Klemm, Anton Nedelin, Nicol\'o Petri, Michela Petrini, Marco Rabbiosi, Henning Samtleben and Alessandro Tomasiello for useful discussions and comments.
Special thanks go to Alberto Zaffaroni for commenting on a draft of this paper and for several illuminating conversations.
We also like to acknowledge the collaboration with Oscar de Felice, Nick Halmagyi and Noppadol Mekareeya on related topics.
SMH is supported in part by INFN. KH is supported in part by the Bulgarian NSF grant DN08/3. AP is supported by the Knut and Alice Wallenberg Foundation under grant Dnr KAW 2015.0083.

\appendix

\section{Supergravity details}
\label{appendix}
In this appendix we elaborate on the supergravity theory and the black hole solutions we consider in the paper. In particular, we focus on the end of the sequence of massive type IIA supergravity truncations in Fig.\;\ref{trunc_fig}, what we call the ``dyonic STU model''. 

\subsection{Dyonic STU model}
We start with the bosonic part of the Lagrangian for the dyonic STU model, following the notation and conventions of the standard reference \cite{Andrianopoli:1996cm},
\begin{align}\label{dyonic Lagrangian}
\begin{split}
\frac{1}{\sqrt{-g}} {\cal L}_{\text{dyonic STU}} &= \frac{R}{2} - V_{g, m} - g_{i \bar{j}} \partial_{\mu} z^i \partial^{\mu} \bar{z}^{\bar{j}} - h_{u v} \nabla_{\mu} q^u \nabla^{\mu} q^v + \frac{1}{4} {\rm I}_{\Lambda \Sigma} H^{\Lambda \mu \nu} H^{\Sigma}{}_{\mu \nu} \\
 &+ \frac14 {\rm R}_{\Lambda \Sigma} H^{\Lambda \mu \nu} * H^{\Sigma}{}_{\mu \nu} - m \frac{\varepsilon^{\mu \nu \rho \sigma}}{4 \sqrt{-g}} B^0{}_{\mu \nu} \partial_{\rho} A_{0 \sigma} - g m \frac{\varepsilon^{\mu \nu \rho \sigma}}{32 \sqrt{-g}} B^0{}_{\mu \nu} B^0{}_{\rho \sigma}\ .
\end{split}
\end{align}
This is supplemented by a fermionic counterpart which we do not present here. It is however instructive to look at the covariant derivative of the gravitino,
\begin{align}\label{gravitino derivative}
	\nabla_{\mu} \psi_{\nu A} = (\partial_{\mu} - \frac14 \omega_{\mu}^{ab} \gamma_{ab} + \frac{i}{2} A_{\mu}) \psi_{\nu A} + (\partial_{\mu} q^u \omega_u{}_A{}^B - \frac{i g}{2} \langle {\cal P}^x, {\cal A}_{\mu} \rangle \sigma^x{}_A{}^B ) \psi_{\nu B}\ .
\end{align}
Many of the above quantities require explanation, and in what follows we will discuss independently several of the sectors of the theory. 
\\ \\
{\bf Gravity multiplet}
The gravity multiplet consists of the graviton $g_{\mu \nu}$, a doublet of gravitini $\psi_{\mu A}$, which transform into each other under the R-symmetry group $\U(1)_{\rm R} \times \SU(2)_{\rm R}$, and a gauge field called the graviphoton with field strength $T_{\mu \nu}$. Due to the presence of three additional vector multiplets in the theory the total number of gauge fields is four, denoted by $A^{\Lambda}_{\mu} \, ,$ $\Lambda \in \{0,1,2,3\}$. The graviphoton field strength is a scalar dependent linear combination of the four field strengths $F^{\Lambda}_{\mu \nu}$. The theory we consider is gauged, meaning that some of the original global symmetries of the theory have been made local.
\\ \\
{\bf Universal hypermultiplet}
An ${\cal N}=2$ hypermultiplet consists of four real scalars $q^u$ and two chiral fermions $\zeta_{\alpha}$ called hyperini. The scalar moduli space is a quaternionic K\"{a}hler manifold with metric $h_{u v} (q)$ and three almost complex structures which further define three quaternionic two-forms that are covariantly constant with respect to an $\SU(2)$ connection $\omega^x$, $x \in \{1,2,3\}$. The particular model that comes from the truncation of ${\cal N}=8$ ISO$(7)$ gauged supergravity has a single hypermutiplet, which universally appears in various string compactifications, hence called the universal hypermultiplet. The moduli space is the coset space $\SU(2,1) / \U(2)$. The metric, written in terms of real coordinates $\{\phi,\, \sigma,\, \zeta,\, \tilde{\zeta} \}$, is  
\begingroup
\renewcommand*{\arraystretch}{1.2}
\begin{equation}
h = 
\begin{pmatrix}
1 & 0 & 0 & 0 \\
0 & \frac{1}{4} \e^{4\phi} & - \frac{1}{8} \e^{4\phi} \tilde{\zeta} & \frac{1}{8} \e^{4\phi} \zeta \\
0 & - \frac{1}{8} \e^{4\phi} \tilde{\zeta} & \frac{1}{4} \e^{2\phi}(1 + \frac{1}{4} \e^{2\phi} \tilde{\zeta}^2) & -\frac{1}{16} \e^{4\phi} \zeta \tilde{\zeta} \\
0 &  \frac{1}{8} \e^{4\phi} \zeta & -\frac{1}{16} \e^{4\phi} \zeta \tilde{\zeta} &  \frac{1}{4} e^{2\phi}(1 + \frac{1}{4} \e^{2\phi} \zeta^2)
\end{pmatrix}.
\end{equation}
\endgroup
The isometry group $\SU(2,1)$ has eight generators; two of these are used for gauging in the model under consideration, generating the group $\mathbb{R} \times \U(1)$. The corresponding Killing vectors are
\begin{align}
	k^{\mathbb{R}} = \partial_{\sigma}\ , \qquad k^{\U(1)} = -\tilde{\zeta} \partial_{\zeta} + \zeta \partial_{\tilde{\zeta}}\ .
\end{align}
These two isometries are gauged by a particular linear combination of the four vector fields in the theory. One defines Killing vectors with index $\Lambda$ corresponding to each of the four gauge fields, such that the hypermultiplet scalar covariant derivative that appears in \eqref{dyonic Lagrangian} reads
\begin{align}
	\nabla_{\mu} q^u \equiv \partial_{\mu} q^u - g \langle {\cal K}^u, {\cal A}_{\mu} \rangle =  \partial_{\mu} q^u - g k^u_{\Lambda} A^{\Lambda}_{\mu} + g k^{u, \Sigma} A_{\Sigma, \mu}\ ,
\end{align}
where $g$ is the gauge coupling constant and the operation $\langle. , .\rangle$ is the symplectic inner product which will be discussed further when we move to the vector multiplet sector. What is important to notice here is that we allow for the hypermultiplet isometries to be gauged not only by the ``ordinary'' electric fields $A^{\Lambda}_{\mu}$ but also by their dual magnetic fields $A_{\Lambda, \mu}$. In the particular model here, the non-vanishing Killing vectors are
\begin{align}\label{app:magnetic killing}
	k_0 = k^{\mathbb{R}}\ , \quad k^0 = c  k^{\mathbb{R}}\ , \quad k_{1,2,3} = k^{\U(1)}\ , 
	\qquad c \equiv \frac{m}{g} \ ,
\end{align}
which means that the magnetic gauge field $A_{0, \mu}$ explicitly appears in the covariant derivative of the scalar $\sigma$ with an effective coupling constant $m$ related to the Romans mass of the massive type IIA supergravity. 

One can also define moment maps (or momentum maps) associated with each isometry on the quaternionic K\"{a}hler manifold. Using the metric and $\SU(2)$ connection on the universal hypermultiplet scalar manifold (see \eg\;appendix D of \cite{Hristov:2012bk}) we find
\begin{align}
P_0 = \begin{pmatrix} 0, & 0, & - \frac{1}{2} \e^{2\phi} \end{pmatrix} ,& \qquad
P^0 = \begin{pmatrix} 0, & 0, &  -\frac{1}{2} c \e^{2\phi} \end{pmatrix} , \qquad \nn \\
P_{1,2,3} = \begin{pmatrix} \tilde{\zeta} \e^{\phi}, & - \zeta \e^{\phi}, &  1 - \frac{1}{4} (\zeta^2 + \tilde{\zeta}^2) \e^{2\phi}\end{pmatrix} ,& \qquad
P^{1,2,3} = \begin{pmatrix} 0, & 0, & 0 \end{pmatrix} .
\end{align} 
These are the moment maps that appear in the gravitino covariant derivative \eqref{gravitino derivative} as a symplectic vector ${\cal P}^x = (P^{x, \Lambda}, P^x_{\Lambda})$. Even in the absence of hypermultiplets the moment maps can be non-zero, signifying that the R-symmetry rotating the gravitini is gauged. 
\\ \\
{\bf STU vector multiplets}
Each ${\cal N} = 2$ vector multiplet consists of one gauge field, a doublet of chiral fermions $\lambda^A$ called gaugini, and a complex scalar field $z$. We already mentioned that the STU model has three vector multiplets and hence three complex scalars $z^i$, labeled by $s$, $t$, and $u$: $z^1 \equiv s$, $z^2 \equiv t$, $z^3 \equiv u$. The scalar manifold is a special K\"{a}hler space whose metric can be derived from a prepotential ${\cal F}$, which for the STU model is,
\begin{equation}\label{app:prepotential}
	{\cal F} = - 2 \sqrt{X^0 X^1 X^2 X^3} \ .
\end{equation}
$X^{\Lambda} = X^{\Lambda} (z^i)$ define the holomorphic sections ${\cal X} \equiv (X^{\Lambda}, F_{\Lambda})$ where
\begin{align}
	F_{\Lambda} \equiv \frac{\partial {\cal F}}{\partial X^{\Lambda}}\ .
\end{align}
${\cal X}$ transforms as a vector under electromagnetic duality or symplectic rotations which leave the solutions of the theory invariant. Other symplectic vectors are the Killing vectors ${\cal K}^u = (k^{u, \Lambda}, k^u_{\Lambda})$, the moment maps ${\cal P}^x = (P^{x, \Lambda}, P^x_{\Lambda})$, the gauge fields ${\cal A}_{\mu} = (A^{\Lambda}_{\mu},A_{\Lambda, \mu})$, and finally the vector of magnetic $p^{\Lambda}$ and electric $e_{\Lambda}$ charges, ${\cal Q} = (p^{\Lambda}, e_{\Lambda})$, giving the name to the duality.

Returning to the holomorphic sections, we pick the standard parameterization
\begin{align}\label{app:scalar parameterization}
(X^0,\, X^1,\, X^2,\, X^3,\, F_0,\, F_1,\, F_2,\, F_3) = (- s t u,\, -s,\, -t,\, -u,\, 1,\, t u,\, s u,\, s t)\ . 
\end{align}
The metric on the moduli space follows from the K\"ahler potential,
\begin{align}
	{\cal K} = - \log (i \langle {\cal X}, \bar{{\cal X}} \rangle) = - \log (i \bar{X}^{\Lambda} F_{\Lambda} - i X^{\Lambda} \bar{F}_{\Lambda}) = - \log (i (s-\bar{s}) (t-\bar{t}) (u-\bar{u}))\ , 
\end{align}
as $g_{i \bar{j}} \equiv \partial_i \partial_{\bar{j}} {\cal K}$ with $\partial_i = \partial/\partial z^i$. We therefore find that $g_{i \bar{j}}$ is diagonal
\begin{align}
	g_{s \bar{s}} = \frac{1}{4 ({\rm Im} (s))^2}\ , \quad g_{t \bar{t}} = \frac{1}{4 ({\rm Im} (t))^2}\ , \quad g_{u \bar{u}} = \frac{1}{4 ({\rm Im} (u))^2}\ .  
\end{align}

Using the K\"ahler potential we introduce the rescaled sections 
\be
{\cal V} = \e^{{\cal K}/2} {\cal X} = (\e^{{\cal K}/2} X^{\Lambda}, \e^{{\cal K}/2} F_{\Lambda}) \equiv (L^{\Lambda}, M_{\Lambda})
\ee
and covariant derivatives 
\be
D_i {\cal V} = (f_i^{\Lambda}, h_{i, \Lambda}) \equiv \e^{{\cal K}/2} \left((\partial_i X^{\Lambda}+X^{\Lambda} \partial_i {\cal K}), (\partial_i F_{\Lambda}+F_{\Lambda} \partial_i {\cal K})\right).
\ee

Moving on to the kinetic terms for the vector fields, the magnetic gauging of the $\mathbb{R}$ isometry in \eqref{app:magnetic killing}, leads to the appearance of the magnetic field $A_{0, \mu}$ in the covariant derivative of the scalar field $\sigma$. Consistency with supersymmetry then requires the introduction of an auxiliary tensor field $B^0_{\mu \nu}$ as derived in \cite{deWit:2005ub,Samtleben:2008pe}. The Lagrangian \eqref{dyonic Lagrangian} therefore contains the modified field strengths
\begin{align}
	H^0_{\mu \nu} \equiv F^0_{\mu \nu} + \frac{1}{2} m B^0_{\mu\nu}\ , \qquad H^{i=1,2,3}_{\mu \nu} \equiv F^i_{\mu \nu}\ ,
\end{align}
where $F^{\Lambda}_{\mu \nu}$ are the field strengths of the electric potentials $A^{\Lambda}_{\mu}$. The kinetic and theta term for the field strengths $H$ involve the scalar-dependent matrices, ${\rm I}_{\Lambda \Sigma} \equiv {\rm Im} ({\cal N})_{\Lambda \Sigma}$ and ${\rm R}_{\Lambda \Sigma} \equiv {\rm Re} ({\cal N})_{\Lambda \Sigma}$. The matrix $\mathcal{N}$ can be computed from the prepotential via \eg\;\cite[(2.6)]{Klemm:2016wng}.
\\ \\
{\bf Scalar potential}
The last part of the Lagrangian \eqref{dyonic Lagrangian} left to discuss is the scalar potential $V_{g, m}$ which depends on the electric and magnetic gauge coupling constants $g$ and $m$ and is given by the general formula
\begin{align}
	V_{g, m} = g^2\left(4 h_{u v} \langle {\cal K}^u, {\cal V}\rangle  \langle {\cal K}^u, \bar{{\cal V}}\rangle  + g^{i \bar{j}} \langle {\cal P}^x, D_i {\cal V}\rangle  \langle {\cal P}^x, \bar{D}_{\bar{j}} \bar{{\cal V}}\rangle  - 3 \langle {\cal P}^x, {\cal V}\rangle \langle {\cal P}^x, \bar{{\cal V}}\rangle \right) \ .
\end{align}
$V_{g, m}$ can be further evaluated explicitly for the dyonic STU model but we will not need its expression. 

The theory is now fully specified by the data of the hypermultiplet moduli space, the vector multiplet moduli space, derived from the prepotential ${\cal F}$ in \eqref{app:prepotential}, and the Killing vectors \eqref{app:magnetic killing} specifying the gauging.
\\ \\
{\bf Tensor fields}
Due to the presence of the auxiliary tensor field $B^0$ (the other auxiliary fields can be immediately decoupled from the theory), there is an additional constraint arising as an equation of motion for $B^0$,
\begin{align}
	G_{\Lambda, \mu\nu} = F_{\Lambda, \mu \nu} + \frac{1}{2} m B^0_{\mu\nu}\ ,
\end{align}
where $G_{\Lambda, \mu\nu}$ is the dual field strength defined by $G_{\Lambda} = (2/\sqrt{-g}) * \delta {\cal L}/\delta F^{\Lambda}$. This leads to
\begin{align}
	G_{\Lambda, \mu\nu} = \frac12 {\rm I}_{\Lambda \Sigma} H^{\Sigma}_{\mu \nu} + \frac{1}{4 \sqrt{-g}} \epsilon_{\mu\nu\rho\sigma} {\rm R}_{\Lambda \Sigma} H^{\Sigma, \rho\sigma}\ .
\end{align}
The appearance of the magnetic gauge field $A_0$ in the Lagrangian leads to the following equation of motion constraining the auxiliary tensor field
\begin{align}\label{app:B0fixing}
	\frac14 \epsilon^{\mu \nu \rho \sigma} \partial_{\mu} B^0_{\nu\rho} = -2 \sqrt{-g} h_{u v} k^{u, 0}\nabla^{\sigma} q^v\ ,
\end{align}
while the rest of the equations of motion are the standard Einstein--Maxwell equations (with sources) and the scalar equations, stemming from \eqref{dyonic Lagrangian}. We discuss these below after specifying the metric ansatz. Note that the BPS conditions together with the Maxwell equations imply the rest of the equations of motion.\\

\subsection{BPS black holes}
In this subsection we write down a black hole ansatz and derive the corresponding BPS equations.

\subsubsection{Black hole ansatz}
We are interested in supersymmetric asymptotically AdS$_4$ black holes, which in \cite{Klemm:2016wng} were considered for general models with hypermultiplets and dyonic gaugings, extending earlier work \cite{Cacciatori:2009iz,DallAgata:2010ejj,Hristov:2010ri,Halmagyi:2013sla,Katmadas:2014faa,Halmagyi:2014qza}. Reviewing these results, we write down the bosonic field ansatz and the final form of the supersymmetry conditions to be solved, which also imply all equations of motion. The metric is given by
\begin{equation}
{\rm d} s^2 = - \e^{2U(r)} {\rm d} t^2 + \e^{-2U(r)} {\rm d} r^2 + \e^{2(\psi(r) - U(r))}{\rm d}\Omega_{\kappa}^2\ ,\label{eq:ansatzmet} 
\end{equation}
where ${\rm d}\Omega_{\kappa}^2={\rm d}\theta^2+f_{\kappa}^2(\theta){\rm d}\varphi^2$ defines the metric on a surface $\Sigma_\fg$ of constant scalar curvature
$2\kappa$, with $\kappa\in\{+1,-1\}$, and
\begin{equation}
f_\kappa(\theta) = \frac{1}{\sqrt{\kappa}} \sin(\sqrt{\kappa}\theta) = 
\left\{\begin{array}{c@{\quad}l} \sin\theta\, & \kappa=+1\,, \\                                             
                                             \sinh\theta\, & \kappa=-1\,. \end{array}\right. 
\end{equation} 
The scalar fields depend only on the radial coordinate $r$, while the electric and magnetic gauge fields $(A^\Lambda$, $A_{\Lambda})$ and the tensor fields $(B^\Lambda$, $B_\Lambda)$ are given by
\begin{equation}
A^\Lambda = A_t^\Lambda {\rm d} t- \kappa p^\Lambda f^\prime_\kappa(\theta){\rm d}\phi\ , \qquad
A_\Lambda = A_{\Lambda t}{\rm d} t - \kappa e_\Lambda f^\prime_\kappa(\theta){\rm d}\phi\ ,
\end{equation}
\begin{equation}
B^\Lambda = 2\kappa p^{\prime\,\Lambda} f^\prime_\kappa(\theta){\rm d} r\wedge{\rm d}\phi\ , \qquad 
B_\Lambda = -2\kappa e^{\prime}_\Lambda f^\prime_\kappa(\theta){\rm d} r\wedge{\rm d}\phi\ .
\end{equation}
In the theory we consider the only relevant tensor field is $B^0$, as the rest can be consistently decoupled.
The magnetic and electric charges $(p^\Lambda, e_\Lambda)$ are defined as
\begin{equation}
p^{\Lambda} \equiv \frac1{\mbox{vol}(\Sigma_\fg)}\int_{\Sigma_\fg} F^{\Lambda}\ , \quad   e_{\Lambda} \equiv \frac1{\mbox{vol}(\Sigma_\fg)}\int_{\Sigma_\fg} G_{\Lambda}\ , \quad
\mbox{vol}(\Sigma_\fg) = \int f_\kappa(\theta){\rm d}\theta\wedge{\rm d}\phi\ .
\label{eq:charges}
\end{equation}
Note that the charges can depend on the radial coordinate in general, since the Maxwell equations are sourced by the hypermultiplet scalars due to the gauging.

\subsubsection{BPS and Maxwell equations}
The above ansatz is subject to a set of conditions required for supersymmetry, which after a number of manipulations can be recast into a set of algebraic and first order differential equations, given by (3.74) in \cite{Klemm:2016wng} in a manifestly symplectic covariant way,
\begin{align}
\begin{split}
\label{app:BHequations}
{\cal E}  &= 0\ , \\
\psi^{\prime} &= -2\kappa \e^{-U} {\rm Im}(\e^{-i\alpha}\mathcal{L})\ , \\
\alpha^{\prime} + A_r &= 2\kappa \e^{-U}{\rm Re}(\e^{-i\alpha} {\cal L})\ , \\
q^{\prime\,u} &= \kappa \e^{-U} h^{uv}{\rm Im}(\e^{-i\alpha}\partial_v {\cal L})\ ,  \\
{\cal Q}^\prime &= -4 \e^{2\psi - 3U} {\cal H}\Omega {\rm Re}(\e^{-i\alpha}{\cal V})\ ,\\
\end{split}
\end{align}
where
\begin{align}
	{\cal E} \equiv 2 \e^{2\psi}\left(\e^{-U} {\rm Im}(\e^{-i\alpha}{\cal V})\right)^{\prime} - \kappa \e^{2(\psi -
U)}\Omega{\cal M} {\cal Q}^x {\cal P}^x + 4 \e^{2\psi-U}(\alpha^{\prime} + A_r){\rm Re}
(\e^{-i\alpha} {\cal V}) +{\cal Q}\ .
\end{align}
As earlier introduced, ${\cal Q} = (p^{\Lambda}, e_{\Lambda})$, ${\cal P}^x = (P^{x, \Lambda}, P^x_{\Lambda})$ and ${\cal K}^u = (k^{u, \Lambda}, k^u_{\Lambda})$. $A_\mu$ is the $\U(1)$ K\"ahler connection, and $\alpha$ an a priori arbitrary phase of the Killing spinor, which depends only on the radial coordinate (derivatives with respect to which are given by primes). Furthermore,
\begin{align}\label{app:quantities}
{\cal Q}^x & \equiv g \langle {\cal P}^x, {\cal Q}\rangle  = g P^x_{\Lambda} p^{\Lambda} - g P^{x, \Lambda} e_{\Lambda}\ , \quad
{\cal W}^x  \equiv g \langle {\cal P}^x, {\cal V}\rangle  = g  P^x_{\Lambda} L^{\Lambda} - g P^{x, \Lambda} M_{\Lambda}\ ,
\nn \\[.2cm]
{\cal Z} & \equiv \langle {\cal Q}, {\cal V}\rangle  = e_{\Lambda} L^{\Lambda} - p^{\Lambda} M_{\Lambda}\ , \quad
{\cal L}  \equiv {\textstyle \sum_x} {\cal Q}^x {\cal W}^x \ , 
\end{align}
and
\begin{equation}
{\cal H} \equiv g^2 ({\cal K}^u)^T h_{u v} {\cal K}^v \ , \quad
{\cal M} =\left(\begin{array}{cc}
 {\rm I} + 
 {\rm R}  {\rm I}^{-1}  {\rm R} & \,\,-  {\rm R} {\rm I} ^{-1} \\
- {\rm I} ^{-1}  {\rm R} &  {\rm I}^{-1} \\
\end{array}\right) \ , \quad \Omega = \left(\begin{array}{cc} 0 & -{\bf 1} \\ {\bf 1} & 0 \end{array}\right) \ .
\end{equation}
 
The above equations are further supplemented by the constraints
\begin{equation}\label{app:constr}
{\cal H} \Omega {\cal Q} = 0 \ , \qquad {\cal K}^u h_{uv} q'^v = 0 \ , \qquad {\cal Q}^x {\cal Q}^x = 1 \ .
\end{equation}

As already noted, in addition the BPS equations, the Maxwell equations need to be imposed. The rest of the equations of motion then follow. The Maxwell equations sourced by the hypermultiplet scalars evaluated on the specified bosonic ansatz lead to \cite{Klemm:2016wng} a pair of coupled first order differential equations
\begin{equation}
	{\cal A}_t' = -\e^{2(U-\psi)} \Omega {\cal M} {\cal Q}\ , \qquad {\cal Q}' = -2 \e^{2 (\psi-2 U)} {\cal H} \Omega {\cal A}_t \ .
\end{equation}
They are immediately satisfied given the fifth row of \eqref{app:BHequations} together with the extra constraint
\begin{equation}
	2 \e^U {\cal H} \Omega {\rm Re} (\e^{-i \alpha} {\cal V}) = {\cal H} \Omega {\cal A}_t\ .
\end{equation}

\subsubsection{Solution to the constraints}
Without making any further assumptions, we can already solve for some of the scalar fields using the constraints \eqref{app:constr} that need to hold everywhere in spacetime. The first equation in \eqref{app:constr} gives
\begin{equation}
	g p^0 - m e_0 = 0 \ , \qquad (\zeta^2+\tilde{\zeta}^2) \sum_{I=1}^3 p^I = 0 \ ,
\end{equation}
while the last one further fixes
\begin{equation}
g\sum_{I=1}^3 p^I = \pm 1 \ .
\end{equation}
Hence,
\begin{equation}
\zeta = \tilde{\zeta} = 0 \ .
\end{equation}
The second equation in \eqref{app:constr} then yields 
\begin{equation}
	\sigma = \text{const.} \, .
\end{equation}
Following the above results, 
\bea\label{app:quantities-simplified}
{\cal Q}^x & = g{\tsumI} p^I \delta^{x, 3}  = \pm 1 \delta^{x, 3} \equiv \lambda \delta^{x, 3} \ , \\[.2cm]
{\cal W}^x & = g \e^{{\cal K}/2}\left[ \tsumI X^I -\tfrac{1}{2} \e^{2 \phi}(X^0 - c F_0) \right] \delta^{x, 3} \ , \\[.2cm]
{\cal Z} & = \e^{{\cal K}/2} \tsumI ( e_I X^I - p^I F_I) + \e^{{\cal K}/2} e_0 (X^0 - c F_0) \ ,  \\[.2cm]
{\cal L} &= \lambda g \e^{{\cal K}/2}\left[ \tsumI X^I -\tfrac{1}{2}\e^{2 \phi}(X^0 - c F_0) \right] ,
\eea
and the only components of the matrix ${\cal H}$ that remain non-vanishing are
\begin{align}
{\cal H}_{0 0} = \frac14 \e^{4 \phi}\ , \quad {\cal H}^{0 0} = \frac14 c^2 \e^{4 \phi}\ , \quad {\cal H}_{0}{}^{0} = {\cal H}^0{}_0 = \frac14 c \e^{4 \phi}\ .
\end{align}
We have already solved for three of the four hypermultiplet scalars, so it is worth writing explicitly the differential equation that determines the remaining scalar $\phi$, coming from the fourth equation of \eqref{app:BHequations}:
\begin{equation}\label{app:phi}
	\phi' = - g \kappa \lambda \e^{{\cal K}/2 - U} {\rm Im} \left(\e^{-i \alpha} (X^0 - c F_0)\right)\ . 	
\end{equation}
The scalar $\phi$ is exactly the source that does not allow the charges $p^0 = c e_0$ to be conserved as it appears in the matrix ${\cal H}$ on the right-hand side of the Maxwell equations, 
\begin{equation}\label{app:charge}
	p'^0 = c e'_0 = -c \e^{2 \psi - 3 U} \e^{4 \phi} {\rm Re}\left(\e^{-i \alpha} (X^0 - c F_0)\right)\ .
\end{equation}
Therefore, the charges $p^0$, $e_0$ cannot actually ``be felt'' by the field theory at the asymptotic AdS$_4$ boundary. 

The equations have been simplified, and are given by the scalar equation \eqref{app:phi}, the Maxwell equation \eqref{app:charge}, and the first three equations in \eqref{app:BHequations}. Note that in the absence of the hypermultiplet equations, \eqref{app:phi} and \eqref{app:charge} are solved trivially, and the remaining equations in \eqref{app:BHequations} can be solved analytically using standard special geometry identities. Here, we are unable to present an analytic solution for the full black hole geometry, exactly due to the complication of solving \eqref{app:phi} and \eqref{app:charge}.
We are however able to present an analytic solution for the two end-points of the black hole geometry, due to the extra condition of the scalars and charges being constant.

Before moving to the ``constant scalars and charges'' case, let us give the relevant components of the matrix ${\cal M}$ which allow us to write down the first equation in \eqref{app:BHequations}:
\bea
{\cal M}^{00} &= -8 \e^{\cal K} |s|^2 |t|^2 |u|^2, \quad {\cal M}^0{}_0 =  -8 \e^{\cal K} {\rm Re} (s)\ {\rm Re} (t)\ {\rm Re} (u) \ , \\
{\cal M}^{01} &=-8 \e^{\cal K} {\rm Re} (t)\ {\rm Re} (u) |s|^2 \ , \\
{\cal M}^{02} &=-8 \e^{\cal K} {\rm Re} (s)\ {\rm Re} (u) |t|^2 \ , \\ 
{\cal M}^{03} &=-8 \e^{\cal K} {\rm Re} (s)\ {\rm Re} (t) |u|^2 \ , \\
{\cal M}^{10} &= {\cal M}^{01} , \quad
{\cal M}^1{}_0 = -8 \e^{\cal K} {\rm Re} (s) \ , \\
{\cal M}^{11} &=-8 \e^{\cal K} |s|^2 \ , \quad 
{\cal M}^{12} =-8 \e^{\cal K} {\rm Re}(s) {\rm Re}(t) \ , \quad {\cal M}^{13} = -8 \e^{\cal K} {\rm Re}(s) {\rm Re}(u) 
\ , \\
{\cal M}^{20} &= {\cal M}^{02} , \quad {\cal M}^2{}_0 = -8 \e^{\cal K} {\rm Re} (t) \ , \\
{\cal M}^{21} &= {\cal M}^{12} \ , \quad
{\cal M}^{22} = -8 \e^{\cal K} |t|^2 \ , \quad 
{\cal M}^{23} = -8 \e^{\cal K} {\rm Re}(t) {\rm Re}(u) \ , \\
{\cal M}^{30} &= {\cal M}^{03} , \quad {\cal M}^3{}_0 = -8 \e^{\cal K} {\rm Re} (u) \ , \\
{\cal M}^{31} &= {\cal M}^{13} \ , \quad
{\cal M}^{32} = {\cal M}^{23} \ , \quad 
{\cal M}^{33} = -8 \e^{\cal K} |u|^2 \ , \\
{\cal M}_{00} &= -8 \e^{\cal K} , \quad {\cal M}_0{}^0 = {\cal M}_1{}^1 = {\cal M}_2{}^2 = {\cal M}_3{}^3 =  -8 \e^{\cal K} {\rm Re} (s)\ {\rm Re} (t)\ {\rm Re} (u)\ , \\
{\cal M}_0{}^{1} &= {\cal M}^1{}_0 \ , \quad  {\cal M}_0{}^{2}= {\cal M}^2{}_0 \ , \quad  {\cal M}_0{}^{3} = {\cal M}^2{}_0 \ , \\
{\cal M}_1{}^0 &= -8 \e^{\mathcal{K}} {\rm Re}(s) |t|^2 |u|^2 \ , \quad 
{\cal M}_2{}^0 =  -8 \e^{\mathcal{K}} {\rm Re}(t) |s|^2 |u|^2 \ , \quad  
{\cal M}_3{}^0 =  -8 \e^{\mathcal{K}} {\rm Re}(u) |s|^2 |t|^2 \ , \\
{\cal M}_1{}^2 &= -8 \e^{\cal K} |t|^2 {\rm Re}(u) \ , \quad  {\cal M}_1{}^3 = -8 \e^{\cal K} |u|^2 {\rm Re} (t)
\ , \\ 
{\cal M}_2{}^1 &= -8 \e^{\cal K} |s|^2 {\rm Re} (u) \ , \quad {\cal M}_2{}^3 = -8 \e^{\cal K} |u|^2 {\rm Re}(s) \ ,  
 \\
{\cal M}_3{}^1 &= -8 \e^{\cal K} |s|^2 {\rm Re} (t) \ , \quad {\cal M}_3{}^2 = -8 \e^{\cal K} |t|^2 {\rm Re} (s) \ .
\eea

\subsubsection{Constant scalars and charges}
The condition that all scalars and charges are constant, 
\begin{equation}
	s'= t'=u' = 0\ , \quad q'^u = 0\ , \quad {\cal Q}' = 0\ ,
\end{equation}
(based on the symmetries of AdS$_4$ and AdS$_2 \times \Sigma_\fg$), upon imposed on \eqref{app:phi}-\eqref{app:charge} yields
\begin{equation}
X^0 - c F_0 = 0 \ . 
\end{equation}
This is a strong constraint on the special K\"ahler manifold, leading to 
\begin{equation}
	s t u = -c \ ,
\end{equation}
and therefore  $(X^1,X^2,X^3,F_1, F_2,F_3) = (c/(t u),-t,-u, t u, -c/t, -c/u)$, which are consistent with the prepotential
\begin{equation}
	{\cal F}^\star = -\frac{3}{2} (-c)^{1/3} (X^1 X^2 X^3)^{2/3}\ .
\end{equation}

\noindent{\bf Asymptotic AdS$_4$}
The constant scalars and charges assumption holds for the AdS$_4$ vacuum, which satisfies the BPS equations asymptotically with
\begin{equation}
U = \log(r/L_{{\rm AdS}_4}), \qquad \psi = \log(r^2/L_{{\rm AdS}_4}),
\end{equation}
and
\begin{equation}
s = t = u = \e^{i \pi/3} c^{1/3}, \qquad \e^{2\phi} = 2 c^{-2/3} \ .
\end{equation}
If substitute the above field configuration in \eqref{app:BHequations}, as $r \rightarrow \infty$, we find 
\begin{equation}
	\alpha =- \frac{\pi}{6} \ , \qquad L_{{\rm AdS}_4} = \frac{1}{g}\ \frac{c^{1/6}}{3^{1/4}}\ , 
\end{equation}
which can be easily seen to solve the second and third equations in \eqref{app:BHequations}. The remaining one, ${\cal E} = 0$, is also asymptotically solved as can be verified by
\begin{equation}
	\frac{2}{L_{{\rm AdS}_4}} {\rm Im} (\e^{-i \alpha} {\cal V}) = - \kappa \Omega {\cal M} {\cal Q}^x {\cal P}^x =3 g c \e^{\cal K}(c^{1/3}, -2 c^{-1/3},-2 c^{-1/3},-2 c^{-1/3}, c^{-2/3}, 1,1,1)\ .
\end{equation}
Note that there is no way of fixing the asymptotic values of the massive vector charges $p^0 = c e_0$, but in the process we have fixed uniquely $\lambda$ to be aligned with $\kappa$ so that
\begin{equation}
	\kappa \lambda = -1 \quad \text{or} \qquad \lambda = -\kappa \ ,
\end{equation}
for a choice of positive electric coupling constant $g > 0$. 
\\ 
\\ 
{\bf Near-horizon geometry}
The near-horizon equations are more involved than the asymptotic ones but we are again in the constant scalar case which guarantees that $s t u = -c$ solving the fourth and fifth equation in \eqref{app:BHequations}. The general near-horizon solution was analyzed in \cite{Guarino:2017pkw} but here we make an inspired ansatz for the scalars in a way that enforces $s t u = -c$:
\begin{equation}
	s = \frac{\e^{i \pi/3} c^{1/3} H^1}{(H^1 H^2 H^3)^{1/3}}\ , \quad t = \frac{\e^{i \pi/3} c^{1/3} H^2}{(H^1 H^2 H^3)^{1/3}}\ , \quad u = \frac{\e^{i \pi/3} c^{1/3} H^3}{(H^1 H^2 H^3)^{1/3}}\ ,
\end{equation}
under the condition that  $H^1 + H^2 + H^3 = 1$. With this ansatz we have imposed equal phases of the three scalars meaning we are killing some of the degrees of freedom, and practically restricting the solution to what we call ``purely magnetic'' solution (see the discussion in the main body of the paper). The metric function ansatz is naturally given by
\begin{equation}
U = \log(r/L_{{\rm AdS}_2}), \qquad \psi = \log(L_{\Sigma_\fg} \cdot r / L_{{\rm AdS}_2}),
\end{equation}
where $L_{{\rm AdS}_2}$ is the radius of AdS$_2$ and $L_{\Sigma_\fg}$ that of the surface $\Sigma_\fg$. With this ansatz we solve the second and third equation in \eqref{app:BHequations} by setting
\begin{equation}
	\alpha = -\frac{\pi}{6}\ , \qquad L_{{\rm AdS}_2} = \frac{\e^{-{\cal K}/2} (H^1 H^2 H^3)^{1/3}}{2 g c^{1/3}}\ .
\end{equation}
The remaining symplectic vector of equations ${\cal E} = 0$ can be solved in several steps. The condition that $p^0 = c e_0$ imposes the constraint that ${\cal E}^0 = c {\cal E}_0$ which leads to
\begin{equation}
	\e^{2 \phi} = \frac{2 c^{-2/3}}{3 (H^1 H^2 H^3)^{1/3}}\ , \qquad p^0 = c e_0 = \frac{g c^{1/3}}{3 \sqrt{3} (H^1 H^2 H^3)^{1/3}} L^2_{\Sigma_\fg}\ , 
\end{equation}
while the electric charges are fixed by the components ${\cal E}_{1,2,3}$ to be
\begin{equation}
	e_1 = e_2 = e_3 = e = - \frac{g}{\sqrt{3}}  L^2_{\Sigma_\fg}\ .
\end{equation}
Note that the electric charges are equal and eventually fixed in terms of the magnetic charges, so they are not independent degrees of freedom. However, from the explicit expression it is clear that the value of $e$ is strictly not allowed to vanish, in accordance with the results in \cite{Guarino:2017eag,Guarino:2017pkw}. Finally, equations ${\cal E}^{1,2,3} = 0$ become
\begin{equation}
	\frac{2 g L^2_{\Sigma_\fg}}{3 \sqrt{3} c^{1/3} (H^1 H^2 H^3)^{2/3}}  = \frac{p^1}{H^1 (3 H^1-2)} = \frac{p^2}{H^2 (3 H^2-2)} = \frac{p^3}{H^3 (3 H^3-2)}\ ,
\end{equation}
which are solved by 
\begin{equation}
	L^2_{\Sigma_\fg} = - \frac{\sqrt{3}}{2 g} c^{1/3} (H^1 H^2 H^3)^{2/3} \sum_{I=1}^{3} \frac{p^I}{H^I} \, ,
\end{equation}
together with
\begin{equation}
3 H^I = 1 \pm \sum_{J, K} \frac{\left| \epsilon_{I J K} \big( p^J - p^K \big) \right|}{2 \sqrt{\big(\sqrt{\Theta} \pm p^I\big)^2 - p^J p^K}} \ ,
\end{equation}
where $\epsilon_{I J K}$ is the Levi--Civita symbol and $\Theta$ is defined in \eqref{theta}.

\bibliographystyle{ytphys}

\bibliography{massiveIIA}{}

\end{document}